\documentclass[fleqn,10pt]{wlscirep}
\usepackage[utf8]{inputenc}
\usepackage[T1]{fontenc}

\newcommand{\HC}[0]{heteroclinic cycles}
\newcommand{\SCPS}[0]{partial synchrony}

\DeclareMathOperator{\md}{d}

\makeatletter
\def\diff{\@ifnextchar[{\@diffwith}{\@diffwithout}}
\def\@diffwith[#1]#2#3{\frac{\md^{#1} #2}{\md #3^{#1}}}
\def\@diffwithout#1#2{\frac{\md #1}{\md #2}}
\def\pdiff{\@ifnextchar[{\@pdiffwith}{\@pdiffwithout}}
\def\@pdiffwith[#1]#2#3{\frac{\partial^{#1} #2}{\partial #3^{#1}}}
\def\@pdiffwithout#1#2{\frac{\partial #1}{\partial #2}}
\makeatother

\usepackage{lineno}

\title{Modeling Nonlinear Oscillator Networks Using Physics-Informed Hybrid Reservoir Computing}

\author[1,*]{Andrew Shannon}
\author[2,+]{Conor Houghton}
\author[2,+]{David Barton}
\author[2,+]{Martin Homer}
\affil[1]{University of Bristol, School of Computer Science, Bristol, BS8 1TH, United Kingdom}
\affil[2]{University of Bristol, School of Engineering Mathematics and Technology, Bristol, BS8 1TH, United Kingdom}

\affil[*]{corresponding author: andrew.shannon@bristol.ac.uk}

\affil[+]{these authors contributed equally to this work}

\begin{abstract}
Surrogate modeling of non-linear oscillator networks remains challenging due to discrepancies between simplified analytical models and real-world complexity. To bridge this gap, we investigate hybrid reservoir computing, combining reservoir computing with “expert” analytical models. Simulating the absence of an exact model, we first test the surrogate models with parameter errors in their expert model. Second, in a residual physics task, we assess their performance when their expert model lacks key non-linear coupling terms present in an extended ground-truth model. We focus on short-term forecasting across diverse dynamical regimes, evaluating the use of these surrogates for control applications. We show that hybrid reservoir computers generally outperform standard reservoir computers and exhibit greater robustness to parameter tuning.~This advantage is less pronounced in the residual physics task. Notably, unlike standard reservoir computers, the performance of the hybrid does not degrade when crossing an observed spectral radius threshold. Furthermore, there is good performance for dynamical regimes not accessible to the expert model, demonstrating the contribution of the reservoir.
\end{abstract}

\begin{document}
\flushbottom

\maketitle
\thispagestyle{empty}

\section*{Introduction}
Networks of oscillators appear widely across engineering, and in both the physical and biological sciences. When the networks are non-linear their dynamical behavior can be complex, displaying synchronization, chaos, and traveling waves \cite{bick2018chaos, dorfler2014synchronization, iatsenko2013stationary}. Creating predictive dynamical models is important in areas such as surrogate modeling of non-linear oscillator networks (NLONs) for smart electrical grid optimization  \cite{dorfler2013synchronization,skardal2015control,liu2022stability}, biological computing \cite{ren2021cardiac}, synthetic biology \cite{elowitz2000synthetic,goldbeter2001simple} and the diagnosis and treatment of neurological disorders such as epilepsy \cite{tsakalis2005control} and Parkinson's disease \cite{lozano2019deep,holt2014origins,arlotti2016adaptive,santaniello2010closed}.
These surrogate models have two broad applications: parameter inference and control. Parameter inference can help identify critical states and associated parameter values within a system, ideally also exposing underlying mechanistic processes. Control using surrogate models aims to exploit system knowledge to improve control performance. Methods such as model predictive control \cite{schwenzer2021review} and model-based reinforcement learning \cite{moerland2023mbrl} are examples.

In this paper, we focus on hybrid reservoir computing (RC), a specific form of physics-informed machine learning (PIML). In particular, with a view to probing its viability for control applications, we investigate how well hybrid RC performs surrogate modeling of NLONs.

Real NLONs are often high dimensional, partially observable, noisy, and involve complex intra-network interactions. As such, it can be extremely difficult to create accurate surrogate models. The classical approach to this task is direct \textit{physics-based} modeling, where mechanisms are pre-ordained and parameters fit to data. Machine learning (ML), or \textit{data-driven} modeling, is an alternative approach which uses fully parameterized models and with parameters updated using learning algorithms. These two contrasting methods confer distinct benefits and drawbacks. 

Physics-based models are physically accurate within the bounds of the assumptions made in their construction. Similarly, within the domain of their training data, data-driven models perform well. Failure when predicting \textit{out-of-domain} is therefore common to the two approaches. ML models can be updated \textit{online} using new data to adapt to new situations, although this is in itself a challenging problem \cite{goring2024out}. Physics-based models are inherently interpretable: each term generally has some understood physical meaning, or represents some physical laws or constraints. On the other hand, ML discards this prior knowledge in favor of complete parameterisation from observed data. When data-driven models have many parameters, they require large amounts of data. Physics-based models tend to have few parameters that need fitting to data, and therefore generally require less. Both methods can be computationally expensive as physics-based models rely on complex numerical schemes, and data-driven models often need to use extensive training algorithms for parameter updates.

PIML is a recently-formulated approach for surrogate modeling and prediction applications \cite{karniadakis2021physics}. It combines both physics-based and data-driven methods, attempting to make use of the best features of each. Its goal is to obtain physically constrained, robust, and interpretable models that capture both expert knowledge of dynamical processes and the information that can be extracted from data obtained from sensing and recording devices. PIML models promise to be more data efficient as they do not require all of the dynamics to be learned from scratch, and they may also facilitate adaptivity through the use of machine learning. PIML-based control for NLONs may thus result in more robust, efficient and accurate controllers that are adaptable and generalisable.

For PIML-based modeling of dynamical systems, it is natural to consider ML components with a time component or sequential nature. For instance, recurrent neural networks (RNNs) and their variants (LSTMs \cite{hochreiter1997long}, GRUs \cite{cho2014properties}) are a common choice \cite{vermaak1998recurrent,schafer2006recurrent,gajamannage2023recurrent}. However, RC \cite{jaeger2001echo,maass2002real} is a particularly promising alternative, due to its simple training procedure. The simplicity of an RC may also offer a unique benefit for PIML parameter inference; while a large RNN can learn, or over learn, a complete model of the data without any input from the physics-based component, an RC has a limited capacity which may force the model to use the physics-based component, making it more interpretable. The small number of parameters used by an RC may also further enhance the low-data requirement conferred by the use of system knowledge. RCs are a restricted form of recurrent neural network (RNN), where learning only takes place in an external readout layer. Sequential data is passed into the reservoir via a fixed, random input weight matrix. An update rule then acts as a discrete non-linear map upon the internal state stored within the activations of the reservoir nodes. The result is a high-dimensional non-linear filter of the incoming data with a fading memory of past states. Scaling the weights of the input matrix and internal connectivity controls the extent to which past-state information is maintained in the hidden state and how much influence is exerted by the input data. To compute an output from the reservoir’s internal state, an output weight matrix is trained, often using regularized linear regression. The readout layer may be trained to perform $n$-step-ahead prediction. When configured to predict the next step in a sequence, the reservoir may be run autoregressively with its output fed back in as the next input instance and thus used for time-series forecasting. This is the format we are considering here: using RCs for the prediction of dynamical system trajectories. 

The main advantage of RCs over more complex RNNs is their ease of initialization and simplicity of training. Good time-series forecasting performance can be achieved using only linear regression, even when chaotic dynamics are present \cite{bollt2021explaining}, and issues such as the vanishing gradient problem are avoided. Since they require only general non-linear high-dimensional filtering of inputs and a fading memory of past inputs, RCs can also be constructed in a wide range of physical substrates \cite{yu2023tapered,hauser2021physical}. 

\begin{figure}
\begin{center}
\includegraphics[]{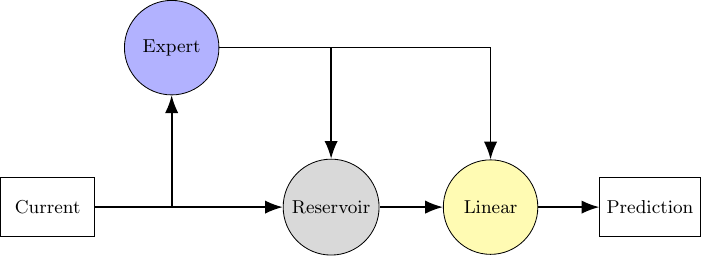}
\end{center}
\caption{A cartoon diagram of the hybrid RC. The current observed state acts as input for both the expert model and the reservoir; the reservoir also receives input from the expert model. The outputs from both the reservoir and expert model form the input to the linear regression layer, whose output maps to a prediction. The only tunable parameters in the model are in the regression layer.}
\label{fig:hrcCartoon}
\end{figure}

Recently, a PIML variant of an echo state network RC was proposed\cite{pathak2018hybrid}. In \textit{hybrid RC} (Fig.~\ref{fig:hrcCartoon}), the prediction from a standard RC is augmented by a single-step integration of an expert ordinary differential equation (ODE) model of the system being predicted. The next step prediction of the ODE model is passed into the reservoir alongside the current state. It is also passed around the reservoir to be considered by the output weight matrix on its own merit, during training and inference. The output weight matrix, still the only trained component, thus aims to combine the augmented reservoir state with the ODE model prediction to most accurately predict the next state. This approach allows the reservoir to compensate for errors in the expert ODE model, and has been shown to result in superior performance when compared to models that use only one of the two components, that is, a standard RC or an expert ODE model.
In particular, the hybrid RC was used to predict the dynamics of the Lorenz and  Kuramoto-Sivashinsky systems when incorporating a model of each system with parameter error. With the correct model structure, the hybrid RC was shown to perform well, better than either a standard RC or ODE model in isolation. The hybrid RC also maintained good performance when, under particular parameter settings, the standard RC and ODE model performed poorly \cite{pathak2018hybrid}.

To investigate the potential of hybrid RCs for the novel example of NLON prediction and control --- where the ground-truth dynamics is more complex, or the expert model is further from the true system than in previous work\cite{pathak2018hybrid} --- we evaluated their performance on two tasks: \textbf{parameter error} and \textbf{residual physics}. 
\begin{itemize}
\item Parameter error, the first, simpler, task tests how well a hybrid RC predicts the trajectory of a network of standard Kuramoto oscillators, when the parameters in the hybrid RC model do not correctly match the parameters of the ground truth model. This follows the previous evaluation but with the Kuramoto oscillator network replacing the Lorenz system \cite{pathak2018hybrid}. The test is run across a range of hyperparameters to assess performance robustness to tuning, and across three qualitatively different dynamical regimes. 
\item Residual physics, our second task, is more challenging. We measure the short term prediction performance of the hybrid RC when the hybrid RC uses a simpler model than the ground truth. This is intended to mimic real-world examples where the complex interactions of an oscillating system are unknown. In these examples a simpler, approximate model, is often used, but even small non-linear terms can quickly make predictions inaccurate. Our interest is in whether the reservoir component of the hybrid RC can compensate for the over-simplification of the model. 
\end{itemize}

In our implementation, the residual physics is an additional higher harmonic in the coupling term for the ground truth Kuramoto-like system: we give the hybrid RC the standard Kuramoto model without this addition. This \textit{bi-harmonic} Kuramoto model \cite{clusella2016minimal} produces behaviors not accessible to the original Kuramoto model. For example, when clustering of the oscillators around a phase occurs in the standard Kuramoto model, there is only one cluster; with an extra harmonic term this need not be true. The residual physics task aims to replicate realistic control scenarios with incomplete knowledge of the system structure. This is an interesting challenge for the hybrid RC since exact knowledge of the ground truth non-linearities has previously been identified as being crucial for Kuramoto oscillator network attractor reconstruction when using the Next Generation Reservoir Computer \cite{gauthier2021next,zhang2023catch}. We run this test across a range of hyper parameters, with four qualitatively different dynamical regimes, using the results to inform a demonstrative grid-search optimization process simulating the development of a surrogate model for control applications. Overall, we are not seeking the best method for modeling NLONs \cite{srinivasan2022parallel,li2024higher}, instead, we look to understand if the benefit of the hybrid approach seen in ~\cite{pathak2018hybrid} carries over to application to NLONs and the more realistic residual physics task.

\section*{Methods}
We present first an overview of the standard and hybrid RCs used in this study, followed by a shared procedure for ground truth data generation and testing in each task. We then describe the specific initialization and training methods, followed by details of the parameter error and residual physics tasks. The details of the hybrid RC and parameter error task largely follow the approach in ~\cite{pathak2018hybrid}. They can therefore be skipped if the reader is familiar with this prior work. The only exception is our addition of the ``Phase component transformation'' to deal with the phase variables.~Our aim is to test hybrid reservoir computing as an approach, rather than to tune a model to a particular application, so we have used as a baseline the parameter settings previously used in ~\cite{pathak2018hybrid}; this isolates the consequence of using the new ground truth system and the new residual physics task from the effect of parameter tuning.

\subsection*{Reservoir Computing}

\begin{figure}[ht]
\centering
\includegraphics[width=0.7817\linewidth]{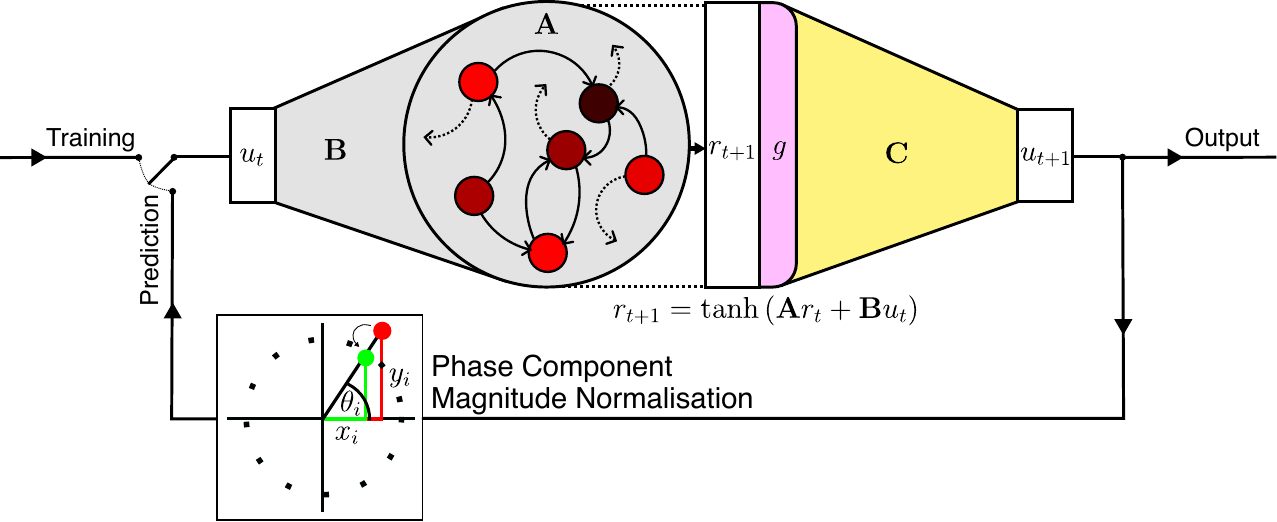}
\caption{The echo state network RC architecture. Input data instance $u_t$ is passed into the reservoir by input weight matrix, $\mathbf{B}$. The internal state of the reservoir $r_t$ is updated according to the update rule, which transforms $r_t$ by the internal weight matrix, $\mathbf{A}$, combines it with the input data, and applies a non-linear transformation to it. The resulting internal state, $r_{t+1}$, is first transformed by the non-linear function, $g$, before being read by the trained output matrix, $\mathbf{C}$, to form the next state prediction $u_{t+1}$. This is then fed back into the reservoir, allowing autoregressive trajectory forecasting. To maintain a magnitude of $1.0$ for the phase components, the output is first transformed into a phase, and back again to be input into the reservoir. The hybrid reservoir may be operated in a \textit{training} or \textit{prediction} mode, for which it must be run in feed-forward and autoregressive fashion respectively.}
\label{fig:SRC_diagram}
\end{figure}

We use the echo state network (ESN) \cite{jaeger2001echo} formulation of RC throughout this work (Fig.~\ref{fig:SRC_diagram}). An ESN comprises $D_r$ nodes with internal states at time $t$ denoted $\mathbf{r_t}\in \mathbb{R}^{D_r\times 1}$. The nodes are coupled together via a weight matrix $\mathbf{A} \in \mathbb{R}^{D_r\times D_r}$. Two weight matrices, $\mathbf{B} \in \mathbb{R}^{D_r\times D_u}$ and $\mathbf{C} \in \mathbb{R}^{D_u\times D_r}$, corresponding to the input weights and the output weights respectively complete the architecture. Input data, $\mathbf{u}_t\in\mathbb{R}^{D_u\times 1}$, is fed into the network via the input weight matrix, $\mathbf{B}$. An update rule operates upon the hidden state and input data to produce the next internal state. For an ESN without a leak term, as used here, the update rule is
\begin{equation} \label{eq:reservoir_update}
    \mathbf{r}_{t+1}=\tanh\left(\mathbf{A}\mathbf{r}_{t}+\mathbf{B}\mathbf{u}_{t}\right).
\end{equation}
Only the output weight matrix, $\mathbf{C}$, is trained, generally via regularized linear regression to produce an output that can take many forms depending on the task. Here, we are focused on next-step prediction for dynamical-systems forecasting, so the output matrix is trained to estimate the next state of the ground truth trajectory. An internal state history matrix $\mathbf{R}=[\mathbf{r}_0,\mathbf{r}_1,\dots,\mathbf{r}_{T}]\in\mathbb{R}^{D_r\times n_T}$ is formed by passing a training trajectory, $\mathbf{u}_T\in\mathbb{R}^{D_u\times n_T}$, into the reservoir in a feed-forward fashion (Fig.~\ref{fig:SRC_diagram} training mode). The linear-regression step optimizes the weights in $\mathbf{C}$ to best form a map between $\mathbf{R}$ and the corresponding next step states of $\mathbf{u}_T$.

To produce a forecast from a given initial condition, an ESN must be run autoregressively, using its output as the next input (Fig~\ref{fig:SRC_diagram} prediction mode). This requires first that the internal reservoir state is synchronized to the start of the trajectory it is forecasting. This \textit{warm-up} phase, with the RC run in feed-forward mode, tends to be quite short, and requires that the reservoir satisfies the echo state property \cite{yildiz2012re}.

A non-linear transformation of the reservoir internal states, $\mathbf{r}_t$, may be applied before the output computation to enrich the representations or break symmetries in the input data \cite{lu2017reservoir}.
\subsection*{Hybrid Reservoir Computing}
\begin{figure}[ht]
\centering
\includegraphics[width=\linewidth]{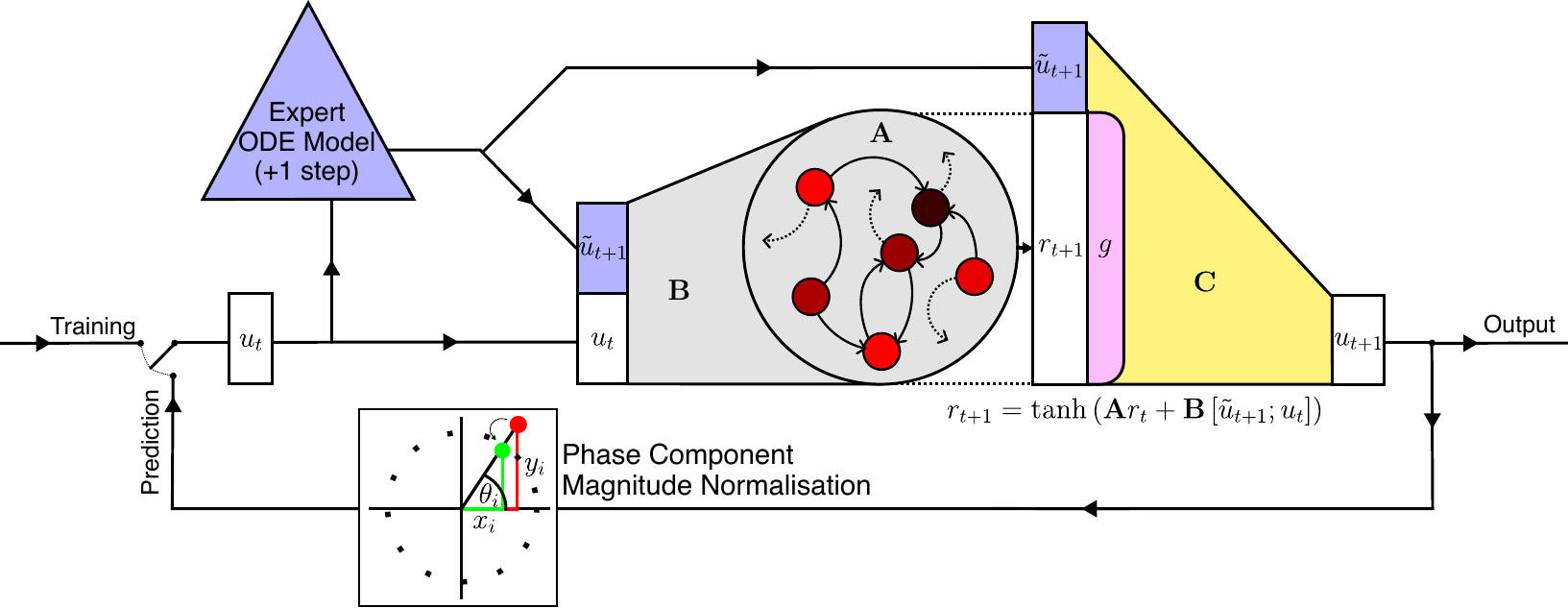}
\caption{The hybrid RC architecture. The incoming state $u_t$ is used as the initial condition of a one step integration of the expert ODE model and then concatenated with the result. This augmented state, $\left[\tilde{u}_{t+1};u_t\right]$, is passed into the reservoir by input matrix, $\mathbf{B}$. The internal state of the reservoir $r_t$ is updated through application of the fixed internal weight matrix, $\mathbf{A}$, addition of the input data, and a $\tanh$ nonlinear transformation, to form $r_{t+1}$. This is transformed by the non-linear function, $g$, before being augmented with the ODE integrator next step prediction $\tilde{u}_{t+1}$. The output matrix, $\mathbf{C}$, trained through linear regression, maps the augmented reservoir state to the next state $u_{t+1}$. To maintain a magnitude of $1.0$ for the phase components, the output is first transformed to phase, and back again to be input into the reservoir. The hybrid reservoir may be operated in a \textit{training} or \textit{prediction} mode, for which it must be run in feed-forward and autoregressive fashion respectively.}
\label{fig:HRC_diagram}
\end{figure}
A hybrid RC \cite{pathak2018hybrid} (Fig.~\ref{fig:HRC_diagram}) is constructed out of a standard echo state network with the addition of an ODE model of the target system (the `expert ODE model', Fig.~\ref{fig:HRC_diagram}, blue triangle). An auxiliary next-step prediction $\tilde{\mathbf{u}}_{t+1}$ is computed from the present state $\mathbf{u}_t$ by a numerical integrator. The integrator’s prediction is then concatenated with the present state, $[\tilde{\mathbf{u}}_{t+1};\mathbf{u}_{t}]$, before being passed into the reservoir as usual via the input weight matrix. The integrator's prediction, $\tilde{\mathbf{u}}_{t+1}$, is also passed around the reservoir and concatenated with the reservoir internal states after the internal update, $[\tilde{\mathbf{u}}_{t+1};\mathbf{r}_{t+1}]$. Because of the concatenation of the usual data instances with the predictions of the ODE model, the size of the input and output weight matrices must change accordingly: $\mathbf{B}\in\mathbb{R}^{D_r\times 2D_u}$, and $\mathbf{C}\in\mathbb{R}^{D_u\times (D_r+D_u)}$. If a non-linear transformation is applied, it is applied to the reservoir internal states, $\mathbf{r}_{t+1}$, only, not the augmented state vector as a whole. The dual use of the integrator's prediction $\tilde{\mathbf{u}}_{t+1}$ is present to maintain the same approach as ~\cite{pathak2018hybrid}, despite it obscuring where the benefit of the hybrid approach originates.

The training procedure is the same as for the standard reservoir, except that the internal-state history matrix includes the auxiliary next-state predictions, $\tilde{\mathbf{u}}_{t+1}$, and therefore $\mathbf{C}$ is being optimized to map from the \textit{augmented} state, $[\tilde{\mathbf{u}}_{t+1};\mathbf{r}_{t+1}]$, to the next state, $\mathbf{u}_{t+1}$. Trajectory forecasting again requires a warm-up phase, and proceeds in the same fashion as the standard ESN method with the addition of the computation of the expert ODE model's auxiliary next state predictions.

\subsection*{Shared Test Procedure}
\begin{figure}[t!]
    \centering
    \begin{minipage}{0.3\textwidth}
        \centering
        \begin{tabular}{|l|l|l|}
            \hline
            Span type & length (steps)\\
            \hline
            Training (T) & $100~\text{s}$ (1000)\\
            \hline
            Training-Test Gap & $100~\text{s}$ (1000)\\
            \hline
            Warm-up & $10~\text{s}$ (100)\\
            \hline
            Test & $250~\text{s}$ (2500)\\
            \hline
            Test-Test gap & $40~\text{s}$ (400)\\
            \hline
        \end{tabular}
    \end{minipage}
    \hfill
    \begin{minipage}{0.60\textwidth}
        \centering
        \includegraphics[width=\textwidth]{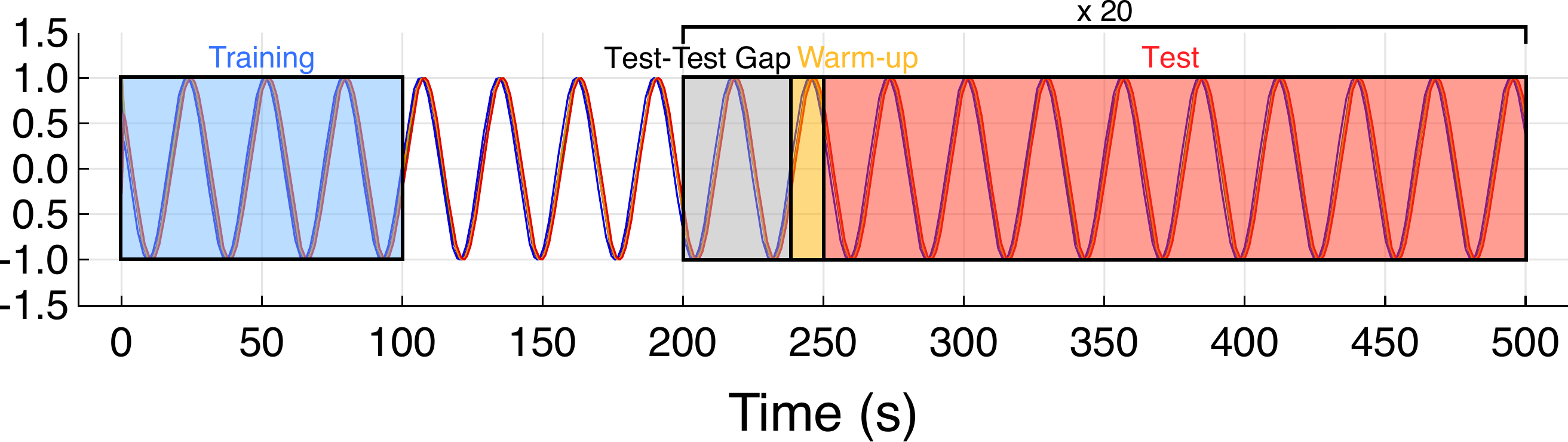}
    \end{minipage}
    \hfill
    \caption{Ground truth trajectory training and test span division. To generate the ground truth data, each dynamical regime was evolved for a long time ($6200~\text{s},~\Delta t=0.1~\text{s}$). The first $1000$ steps were used for training. A gap of $1000$ steps is followed by $20$ warm-up and test segments each composed of $100$ and $2500$ steps respectively, with $400$ step gaps between them.
    \label{fig:trajectory_spans}}
\end{figure}

We used a shared method across each task in this study. In each task, there is a distinct ground truth dynamical system that we are trying to predict. To produce the training, warm-up, and test data for each task, we evolved the ground truth system for a long time and segmented the resulting trajectory. The training, warm-up and test trajectory lengths were chosen according to previous work presenting the hybrid RC architecture \cite{pathak2018hybrid} to enable comparison with the performance achieved there. Figure \ref{fig:trajectory_spans} details the lengths of each stage, and shows schematically the organization of the stages. The long trajectory was divided such that the start is the training span. After a gap, $20$ disjoint warm-up/test spans follow. In every test, regardless of the performance evaluation metric, we initialized $40$ instantiations of the reservoir (standard or hybrid) and trained them on the appropriate training span. Each test then consisted of synchronizing the reservoirs to the initial condition of the test span by feeding in the corresponding warm-up span, and then running the reservoirs autoregressively to produce forecasts of the same length as the test span. As a control, $40$ instantiations of the hybrid RC's expert ODE model (each with sampled parameter error) were produced and integrated with initial conditions matching the test trajectories.

\subsection*{Model Initialization and Training}
\subsubsection*{Standard Reservoirs}
To initialize the standard and hybrid RCs we used the same method as in previous work\cite{pathak2018hybrid}. In particular, the internal connectivity matrix of each reservoir, $\mathbf{A}$, was constructed as an Erd\H{o}s-R\'{e}nyi random graph with mean degree $\left<d\right>$, corresponding to an edge probability of $\left<d\right> /D_r$. The weights were initialized uniformly randomly between $-1$ and $1$, and then scaled such that the spectral radius was set to a desired value. The spectral radius of a matrix is the magnitude of its largest eigenvalue.
The input weight matrix, $\mathbf{B}$, was initialized with a single non-zero element per row, corresponding to one connection per reservoir node. Each of these connections was assigned uniformly randomly to one of the input dimensions. The weight of each connection was drawn from a uniform distribution on a range set by the input scaling hyper-parameter: for all $i\in\{1,\dots,D_r\}$,
\begin{align}
j&\sim\text{Uniform}\{1,\dots,D_u\},\cr
b_{ij}&\sim \text{Uniform}(-\text{input scaling}, \text{input scaling}),\cr
b_{ik}&=0\;\text{for}\;k\not= j.
\end{align}
The spectral radius and input scaling were set to 0.4 and 0.15 respectively as baseline values as in previous work\cite{pathak2018hybrid}. Although restricting the input to one per node may not result in optimum performance due to a lack of mixing at the present step, this was done to match the approach in ~\cite{pathak2018hybrid} to allow easy comparison of the results.

We used the non-linear transformation $g$, applied to the reservoir internal states, $\mathbf{r}_t$, that was used previously\cite{pathak2018hybrid}. The output of the reservoir, $\mathbf{u}_{t+1}$, was thus computed as
\begin{equation} \label{eq:reservoir_output}
    \mathbf{u}_{t+1}=\mathbf{C}g\left(\mathbf{r}_{t+1}\right),
\end{equation}
with the non-linear transformation $g:\mathbb{R}^{D_r}\mapsto\mathbb{R}^{D_r}$ defined such that its $i$-component, $g_i$, is given for all $i\in\{1,\ldots,D_r\}$ by
\begin{equation} \label{eq:reservoir_NLAT}
   g_i(\mathbf{r}) = \begin{cases}
      r_i, & i=\operatorname{odd},\\
      r_i^2, & i=\operatorname{even}.
   \end{cases} 
\end{equation}
The output weight matrix, $\mathbf{C}$, was trained via regularized linear regression to perform next step prediction. An internal state history was formed from the reservoir node activations that were obtained upon processing a training trajectory. $g$ was first applied column-wise to the state history matrix $\mathbf{R}=[\mathbf{r}_0,\mathbf{r}_1,\dots,\mathbf{r}_T]$ such that $\hat{\mathbf{R}}=g_{col}(\mathbf{R})=[g(\mathbf{r}_1),g(\mathbf{r}_2),\dots,g(\mathbf{r}_T)]$. The output weight matrix, $\mathbf{C}$, was then computed using the ridge regression equation
\begin{equation}\label{eq:ridge_regression}    \mathbf{C}=\left(\hat{\mathbf{R}}^{T}\hat{\mathbf{R}}+\beta\mathbf{I}\right)^{-1}\hat{\mathbf{R}}^{T}\mathbf{U}^+,
\end{equation}
where $\mathbf{U}^+\in\mathbb{R}^{D_u\times n_T}$ is the target state matrix, comprising the states one step ahead of the training trajectory states, and $\beta$ is the regularization parameter. $\mathbf{X}^T$ indicates the transpose of $\mathbf{X}$ and $\mathbf{X}^{-1}$ the inverse. $\mathbf{I}$ is the $D_r \times D_r$ identity matrix.
\subsubsection*{Hybrid Reservoirs}
For the hybrid RC, the internal reservoir weight matrix, $\mathbf{A}$, was constructed in the same way as for the standard reservoir. The initialization of the input weight matrix, $\mathbf{B}$, differed from the standard reservoir implementation. A hyper-parameter, the Knowledge Ratio, was used to determine the fraction of connections to the expert ODE model output states. For instance, with a knowledge ratio of $0.3$, $30\%$ of the connections are from states within the expert ODE model output, and $70\%$ are from the incoming data. This was achieved by connecting each reservoir node to either the expert model or the input data, with probability $KR\in[0,1]$ of connection to the expert model. Again, only one connection per reservoir node to the input states/ODE output was formed and the weights were drawn from a uniform distribution on a range set by the input scaling parameter, so, for all $i\in\{1,\dots,D_r\}$ we set
\begin{align}
    j&\sim\begin{cases}
        \text{Uniform} \left\{1,\dots,D_u \right\} \text{, with probability }KR,\\
        \text{Uniform} \left\{D_u+1,\dots,2\times D_u \right\} \text{, with probability }1-KR.
    \end{cases}\cr
    b_{ij}&\sim \text{Uniform}(-\text{input scaling}, \text{input scaling}),\cr
    b_{ik}&=0\;\text{for}\;k\not= j.
\end{align}
A knowledge ratio of $0.5$ was used as a baseline. A sweep from $0.1$ to $0.9$ knowledge ratio on the parameter error task is presented in the supplementary information (Supplementary Fig.~S6).

The output weight matrix, $\mathbf{C}$, was again trained using regularized linear regression. However, in the hybrid case, the augmented state history matrix $\tilde{\mathbf{R}}\in\mathbb{R}^{(D_u+D_r)\times n_T}$ is formed by concatenation of the auxiliary next step predictions with the reservoir node activations, $\tilde{\mathbf{R}}=[\tilde{\mathbf{U}};\mathbf{R}]$. $\tilde{\mathbf{U}}\in\mathbb{R}^{D_u\times n_T}$ contains the auxiliary next step predictions of the expert ODE model corresponding to each instance of the training data. Each column of $\mathbf{R}$ is thus $[\tilde{\mathbf{u}}_{t+1};\mathbf{r}_{t+1}]$. The activations $\mathbf{r}_{t+1}$ are, in turn, produced by processing the training data state, $\mathbf{u}_t$, \textit{and} the corresponding auxiliary next step prediction, $\tilde{\mathbf{u}}_{t+1}$. Only the reservoir internal states, $\mathbf{r}_{t+1}$, within $\tilde{\mathbf{R}}$ were transformed by $g$ prior to computation of the output matrix, $\mathbf{C}$, or any output states during prediction.

\subsubsection*{Phase Component Transformation}
Dynamical systems models of NLONs are often composed of phase variables, one for each oscillator in the network. As phase is bounded within $[-\pi,\pi]$, representing the angle around the unit circle, a phase variable trajectory can contain discontinuous jumps. To avoid this, we project the phase variables onto the $x$ and $y$ axes, forming continuous $x$ and $y$ \textit{phase components}, which are then processed by the reservoir. To conserve the magnitude of the phase variables upon each iteration of the reservoir (standard or hybrid), we projected the phase components to phases, and back to components again after every step (Fig.~\ref{fig:SRC_diagram} and Fig.~\ref{fig:HRC_diagram}, phase component magnitude normalization).

\subsection*{Parameter Error Task}
In the parameter error task, we were aiming to evaluate the performance of the hybrid RC when its expert model differs from the ground truth only through errors in its parameters. This was the case considered previously\cite{pathak2018hybrid}. We conducted this test to find out if the hybrid RC could provide a forecasting performance improvement when predicting NLONs, as it did for the Lorenz and Kuramoto-Sivashinsky systems. We carried out a parameter sweep over a range of hyperparameter settings to test the robustness of any improvement over the standard RC. 

\subsubsection*{Ground Truth}
The Kuramoto model is a canonical model of a system of coupled non-linear oscillators\cite{strogatz2000kuramoto}. We use it as the ground truth in the parameter error task, and as the expert ODE model of the hybrid RC throughout this study. In particular, we used an all-to-all coupled network of Kuramoto oscillators. This model can demonstrate synchronous, asynchronous, and chaotic behavior. The fixed reference frame form of the model is 
\begin{equation}\label{eq:Kuramoto}
    \diff{\theta_i}{t}=\omega_i+\frac{K}{N}\sum^{N}_{j=1}\sin\left(\theta_j-\theta_i\right),
\end{equation}
where $\theta_i(t)$ are the individual oscillator phases, $\omega_i$ are the oscillator natural frequencies, $K$ is the coupling strength, and $N$ is the number of oscillators.

The different dynamical states of equation (\ref{eq:Kuramoto}) are achieved by varying the global coupling strength $K$ and the natural frequencies $\omega_i$. For large enough $K$, and a tight enough distribution of $\omega_i$, the oscillators will synchronize, or demonstrate phase locking. When the coupling strength is low, or the frequencies far apart, the model will display asynchronous behavior \cite{strogatz2000kuramoto}.

\begin{figure}[ht]
\centering
\includegraphics[width=\linewidth]{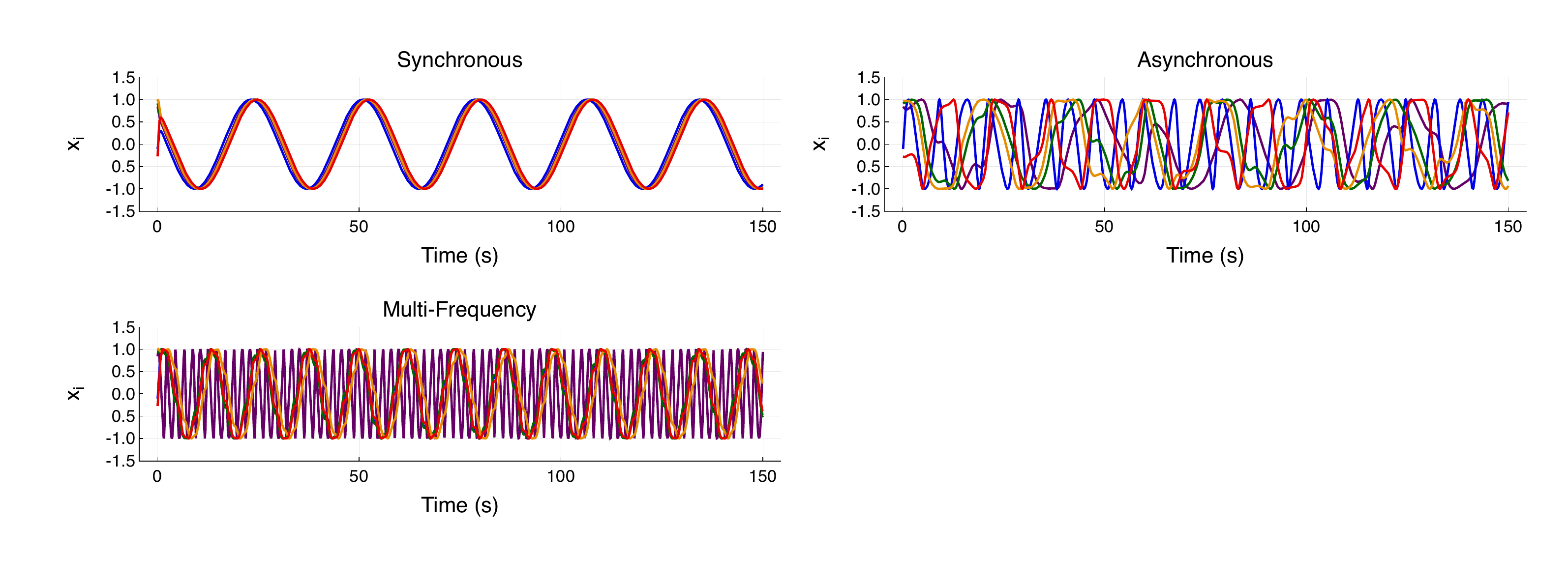}
\caption{Dynamical regimes of the standard Kuramoto oscillator network for the parameter error task. Parameters as in Table \ref{tab:KuramotoParams}. On a scale determined by the particular coupling strength, the asynchronous frequencies are far apart, while the synchronous frequencies are similar; in the multi-frequency regime, one oscillator has a far higher frequency than the other four. Only the $x$ phase component is displayed for each of the five oscillators. }
\label{fig:Kuramoto_regimes}
\end{figure}

We explored three qualitatively distinct dynamical regimes of the Kuramoto model in the parameter error task. They were: a fully synchronized regime, an asynchronous regime, and a multi-frequency regime that has timescale separation between the oscillators but otherwise demonstrates phase locking (Fig.~\ref{fig:Kuramoto_regimes}).

To obtain each regime, the natural frequencies and coupling strengths were varied, with the natural frequencies being drawn from a uniform distribution between $-1$ and $1$. For the multi-frequency regime, the high frequency oscillator's natural frequency was set to $z (3.0 + w)$ where $w$ is a uniformly distributed random number between $0$ and $1$ and $z$ is equal to $-1$ or $+1$ with equal probability. Three realizations of each regime were produced to use as ground truth in the shared test procedure. The parameters for each regime are detailed in Table \ref{tab:KuramotoParams}. 

\begin{table}[ht]
\centering
\begin{tabular}{|l|l|l|l|}
\hline
Regime & $N$ & $\omega_{\;i}$ & $K$\\
\hline
Synchrony & $5$ & $\omega_i \sim \text{Uniform}\left(-1,1\right)$, $i=1\dots5$ & $4.0$ \\
\hline
Asynchrony & $5$ & $\omega_i \sim \text{Uniform}\left(-1,1\right)$, $i=1\dots5$ & $1.0$ \\
\hline
Multi-Frequency & $5$ & $\omega_i \sim \text{Uniform}\left(-1,1\right)$, $i=1\dots4$& $2.0$ \\
 & & $\omega_5=z(3.0 + w)$&  \\
 & & $w\sim \text{Uniform}(0,1)$&  \\
  & & $z\sim \text{Rademacher}()$&  \\
\hline
\end{tabular}
\caption{\label{tab:KuramotoParams}Standard Kuramoto model parameters for each dynamical regime.}
\end{table}

As the Kuramoto model was also being used within the hybrid RC, and therefore needed to be integrated upon every iteration, we used a transformed version of the system of equations to coincide with the $x$, $y$ phase component projection that the reservoirs were using.
Each of the oscillator phase variables $\theta_i$ was transformed into phase components, x$_i=\cos\left(\theta_i\right)$ and $y_i=\sin\left(\theta_i\right)$.
Under this transformation, the Kuramoto model is described by a pair of ODEs for each oscillator, one for each phase component:
\begin{align}
    \diff{x_i}{t}&=-\omega_iy_i-\frac{Ky_i}{N}\sum^{N}_{j=1}\left(y_jx_i-x_jy_i\right),\cr
    \diff{y_i}{t}&=\omega_ix_i+\frac{Kx_i}{N}\sum^{N}_{j=1}\left(y_jx_i-x_jy_i\right).
\end{align}
\subsubsection*{Models}
The standard and hybrid RC's were initialized with a set of baseline parameters and whichever parameter was being varied was set according to the parameter sweep being conducted (Table~\ref{tab:param_ranges_1}). For the parameter error task, the hybrid RC's expert ODE model, and therefore the control ODE model, was the same as the ground truth Kuramoto model. To introduce the parameter error, we added multiplicative error to the coupling strength and to each of the natural frequencies:
\begin{equation}
   p \leftarrow (1 + \xi)p, 
\end{equation}
with $p$ standing in for the parameters and $\xi$ sampled independently for each parameter from a normal distribution, $N(0,\sigma^2_K)$ for coupling strengths and $N(0,\sigma^2_\omega)$ for natural frequencies. The standard deviations $\sigma_K,\sigma_\omega$ were set according to the relevant parameter sweep (baseline 0.05, other settings are explored in the supplementary information, Supplementary Fig.~S1 to S4). The errors were sampled independently for each instantiation.
\subsubsection*{Evaluation Metric}
To evaluate the quality of the trajectory predictions in the parameter error task, we used the normalized mean square error (NMSE)
\begin{equation}
    \text{NMSE}\left(t\right)=\frac{\|\mathbf{u}\left(t\right)-\mathbf{u}^*\left(t\right)\|}{\left<\|\mathbf{u}\left(t\right)\|^2\right>^\frac{1}{2}},
\end{equation}
where $\mathbf{u}^*(t)$ is the prediction and $\mathbf{u}(t)$ the ground truth.
This metric was used in previous work\cite{pathak2018hybrid} to evaluate the valid time metric. For each test, the mean NMSE across the entire trajectory was computed, capturing the overall,~long-term, agreement between the forecast and the ground truth.
\subsubsection*{Parameter Sweep}
A range of hyper parameters were investigated in the parameter error task (Table~\ref{tab:param_ranges_1}). Only the effect of the input scaling and spectral radius parameters are shown here; other results are presented in the supplementary information (Supplementary Fig~S1 to S7). To ensure that a wide range of model instantiations were tested, a random seed dependent on the parameter sweep index was used in the initialization.
\begin{table}[t!]
\centering
\begin{tabular}{|l|l|l|}
\hline
Parameter & Baseline & Range \\
\hline
$D_r$ & 300 & [50, 1000] \\
\hline
Spectral Radius & 0.4 & [0.1, 2.0] \\
\hline
Input Scaling & 0.15 & [0.05, 2.00] \\
\hline
$\sigma_K$ & 0.05 & [0.004, 0.480] \\
\hline
$\sigma_\omega$ & 0.05 & [0.004, 0.480] \\
\hline
Knowledge Ratio & 0.5 & [0.05, 1.00] \\
\hline
Regularization Strength & $10^{-6}$ & Baseline \\
\hline
\end{tabular}
\caption{\label{tab:param_ranges_1}Parameter baselines and ranges for the parameter error task. The parameter sweep modifies individual parameters whilst holding the rest at the baseline setting.}
\end{table}
\subsection*{Residual Physics Task}
In the residual physics task, we evaluated the ability of the hybrid RC to forecast trajectories of a ground truth whose dynamical equations are different from the expert ODE model. To do this we kept the expert ODE model the same as in the parameter error task, i.e., the standard Kuramoto oscillator network model, and used an extension with additional coupling terms to produce a ground truth with new dynamical regimes not accessible to the standard Kuramoto model. In this task, we again evaluated the performance of the standard and hybrid RCs with parameter sweeps, and compared the performance of both RCs to the hybrid RC's expert ODE model alone. Subsequently, we conducted a grid search to investigate the potential of an optimized hybrid RC in comparison to an optimized standard RC.
\subsubsection*{Ground Truth}
We used an extended, bi-harmonic version of the Kuramoto oscillator network model that includes an extra harmonic in the nonlinear coupling term\cite{clusella2016minimal}, given by
\begin{equation}\label{eq:extended_Kuramoto}
    \diff{\theta_i}{t}=\omega_i+\frac{K}{N}\sum^{N}_{j=1}\left[\sin\left(\theta_j-\theta_i+\gamma_1\right)+a\sin\left(2(\theta_j-\theta_i)+\gamma_2\right)\right].
\end{equation}
This introduces a structural difference between the ground truth and the hybrid reservoir’s model; the extra harmonic allows for more complex dynamical regimes, including heteroclinic cycles and self-consistent partial synchrony\cite{clusella2016minimal}. The regimes can be elicited by appropriate choice of the coupling phase shifts $\gamma_1$ and $\gamma_2$ and the scaling of the second harmonic $a$.%
\begin{figure}[ht]
\centering
\includegraphics[width=\linewidth]{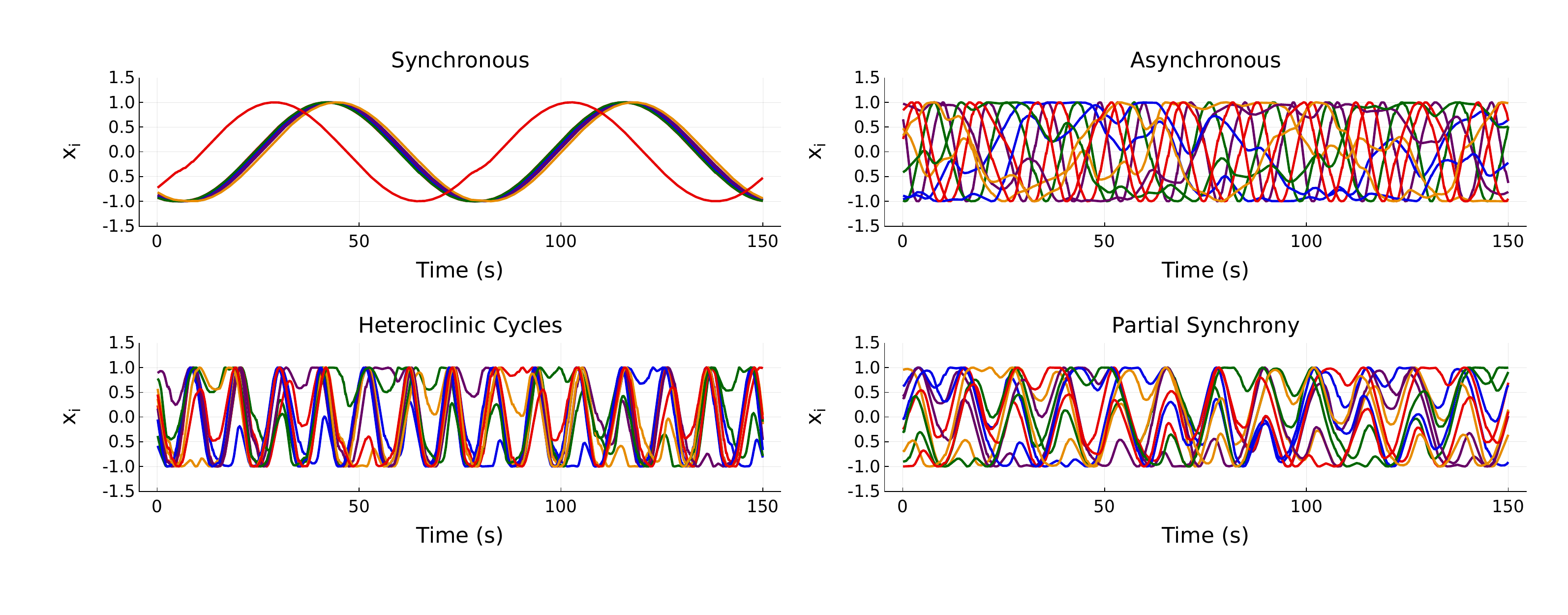}
\caption{Dynamical regimes of the bi-harmonic Kuramoto oscillator network with an extra harmonic in the coupling term, for the residual physics task. Parameters as in Table \ref{tab:ExtKuramotoParams}. Only the $x$ phase component for each oscillator is displayed.}
\label{fig:ExtKuramoto_regimes}
\end{figure}
As with the parameter-error test, we used a set of qualitatively different dynamical regimes to evaluate the performance of the hybrid RC on this task. Four cases were chosen: synchronous, where oscillators are phase locked into one or more clusters, asynchronous, where no phase locking occurs, heteroclinic cycles, where partially synchronized clusters exist and individual oscillators travel between them sporadically, and self-consistent partial synchrony, where a consistent but rotating phase distribution is maintained without full synchronization (Fig.~\ref{fig:ExtKuramoto_regimes}). For brevity the self-consistent partial synchrony regime will be referred to as `partial synchrony' in what follows. We were particularly interested in the \HC~and \SCPS~regimes. Since the standard Kuramoto model does not demonstrate these behaviors,  the hybrid reservoir must learn the structural difference between its expert ODE model and the ground truth. For this task, we sampled the natural frequencies $\omega_i$ from a Lorentzian distribution, $\omega_i\sim\text{Cauchy}(\mu, \Delta\omega)$, where $\mu$ is the center of the distribution, and $\Delta\omega$ is the width. The parameters for each regime \cite{clusella2016minimal} are shown in Table~\ref{tab:ExtKuramotoParams}.

\begin{table}[ht]
\centering
\begin{tabular}{|l|l|l|l|l|l|l|l|l|}
\hline
Regime & $N$ & $\mu$ & $\Delta\omega$ & $K$ & $\gamma_1$ & $\gamma_2$ & $a$\\
\hline
Synchrony & $10$ & $0.0$ & $0.01$ & $1.0$ & $2\pi$ & $\pi$ & $0.2$\\
\hline
Asynchrony & $10$ & $0.0$ & $0.05$ & $5.0$ & $\pi$ & $\pi$ & $0.2$\\
\hline
Heteroclinic Cycles & $10$ & $0.0$ & $0.01$ & $1.0$ & $1.3$ & $\pi$ & $0.2$\\
\hline
Partial Synchrony & $10$ & $0.0$ & $0.01$ & $1.0$ & $1.5$ & $\pi$ & $0.2$\\
\hline
\end{tabular}
\caption{\label{tab:ExtKuramotoParams}Bi-harmonic Kuramoto model parameters for each dynamical regime. Varying the first harmonic phase shift, $\gamma_1$, produces the four different behaviors.}
\end{table}

\subsubsection*{Models}
The standard and hybrid RC's were again initialized using the baseline parameters, and whichever parameters were being varied were set appropriately depending on the parameter sweep or grid search parameter set (Table \ref{tab:param_ranges_2}, Fig.~\ref{fig:grid_search_cube}). For the residual physics task the standard Kuramoto model is used as the hybrid RC's expert ODE model, and therefore the control ODE model. As such, neither model was the same as the ground truth system. We included parameter error to the coupling strength and to each of the natural frequencies in the same fashion as in the parameter error task. 

\subsubsection*{Evaluation Metric}
For the residual physics task, we focus on the application of hybrid RCs to control, and therefore used the valid-time metric to assess short-term prediction quality in both the parameter sweep and the grid search. The valid time $t^*$ is the time during which the predicted trajectory has a NMSE less than some threshold $\epsilon$ in NMSE
\begin{equation}
t^*=\text{max}\left\{t:\text{NMSE}(\tau)\leq\epsilon, \forall \tau\leq t\right\}.
\end{equation}
Throughout this task, we used $\epsilon=0.4$.~This metric only reports the short term accuracy, and therefore does ~not provide information on the long term prediction quality of the models. For our application focus, where short time horizons are often all that is required, this provides a sufficient view of the performance.
\subsubsection*{Parameter Sweep}
We used the same method for the residual physics task that we did previously, for the parameter error task. However, here we varied a different set of parameters. Specifically, we varied the spectral radius, input scaling, regularization strength, and reservoir size (Table~\ref{tab:param_ranges_2}).%
\begin{table}[t!]
\centering
\begin{tabular}{|l|l|l|}
\hline
Parameter & Baseline & Range \\
\hline
$D_r$ & 300 & [50, 1000] \\
\hline
Spectral Radius & 0.4 & [0.0, 2.0] \\
\hline
Input Scaling & 0.15 & [0.1, 2.0] \\
\hline
$\sigma_K$ & 0.05 & Baseline \\
\hline
$\sigma_\omega$ & 0.05 & Baseline \\
\hline
Knowledge Ratio & 0.5 & Baseline \\
\hline
Regularization Strength & $10^{-4}$ & [$10^{-8}$, 0.5] \\
\hline
\end{tabular}
\caption{\label{tab:param_ranges_2}Parameter baselines and ranges for the residual physics task. The parameter sweep modifies individual parameters whilst holding the rest at the baseline setting.}
\end{table}
\subsubsection*{Grid Search}

\begin{figure}[t!]
    \centering
    \begin{minipage}{0.5\textwidth}
        \centering
        \begin{tabular}{|l|l|l|l|l|}
        \hline
        Parameter Set & A & B & C & D \\
        \hline
        Regularization & $0.0001$ & $0.1$ & $0.0001$ & $0.1$ \\
        Spectral Radius & $0.1$ & $0.1$ & $2.0$ & $2.0$ \\
        Input Scaling & $0.05$ & $0.05$ & $0.05$ & $0.05$ \\
        \hline
        Parameter Set & E & F & G & H \\
        \hline
        Regularization & $0.0001$ & $0.1$ & $0.0001$ & $0.1$ \\
        Spectral Radius & $0.1$ & $0.1$ & $2.0$ & $2.0$ \\
        Input Scaling & $0.20$ & $0.20$ & $0.20$ & $0.20$ \\
        \hline
        \end{tabular}
    \end{minipage}
    \begin{minipage}{0.30\textwidth}
        \centering
        \includegraphics[width=\textwidth]{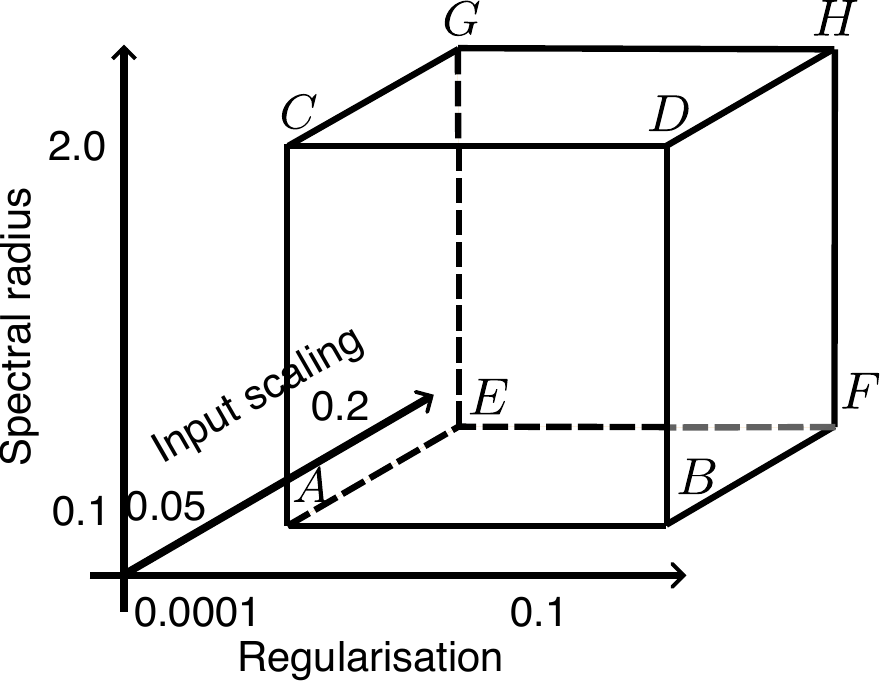}
    \end{minipage}%
    \hfill
    \caption{Grid search parameter combinations for the residual physics task. All combinations of high and low states bounding the optimal region found in the parameter sweep of the regularization, spectral radius, and input scaling parameters were tested. All other parameters were held at baseline (Table \ref{tab:param_ranges_2}). The parameter sets correspond to the corners of the cube in parameter space on the right. Presentation of the results of the grid search follow the parameter sets in alphabetical order, labeled \textbf{A} to \textbf{H}. First at low input scaling, from the bottom left corner (low regularization, spectral radius, and input scaling) to the top, left to right (\textbf{A} to \textbf{D}), and then again at high input scaling, bottom left to the top, left to right (\textbf{E} to \textbf{H}).}
    \label{fig:grid_search_cube}
\end{figure}

To explore the potential of `optimally' tuned standard and hybrid RCs, a grid search was conducted in the residual physics task. We consider here only parameter tuning of the hybrid and standard RCs to maintain a consistent architecture and approach to ~\cite{pathak2018hybrid}. Prior works have explored  optimization of the particular combination of model-based and data-driven components in the hybrid RC \cite{duncan2023optimizing}. High and low parameter settings for the input scaling, spectral radius, and regularization strength were chosen from the results of the parameter sweeps. In particular, these were selected to approximately range the optimum region identified for each parameter, across both the standard and hybrid RC results. A set of all of the combinations of these parameters was produced, labeled from \textbf{A} to \textbf{H}, at each vertex of a cube in parameter space (Fig.~\ref{fig:grid_search_cube}). For each parameter combination, we followed the shared evaluation method as for both of the parameter sweeps. $40$ instantiations of standard and hybrid reservoirs were produced and then trained and tested on ground truth data from three of the four regimes in the residual physics task. These were the synchronous, heteroclinic cycles, and partial synchrony regimes.

\section*{Results}
In this section we demonstrate the capability of the hybrid RC to predict NLONs in various circumstances. We first present results from the parameter error task. Across three qualitatively different dynamical regimes, we compare the mean performance of multiple standard and hybrid RCs. As a control, we test the hybrid RC's expert model, the \textit{base ODE model}, which is also subject to parameter error. We show parameter sweeps of spectral radius and input scaling. Second, we give the results of the residual physics task,  where the hybrid RC is given a reduced form of the ground truth bi-harmonic Kuramoto model. Parameter sweeps are presented for four dynamical regimes. First, we test synchronous and asynchronous regimes. These were present in the standard Kuramoto model, however the bi-harmonic synchronous regime can support multiple clusters. The other two regimes are heteroclinic cycles, and \SCPS; both are inaccessible to the standard Kuramoto model. We test the effects of spectral radius, input scaling, regularization strength and reservoir size. Short-term prediction quality is evaluated to investigate viability for control applications. Finally, the results of a grid search optimization process across three parameters are presented, illustrating the potential of a tuned hybrid RC.
\subsection*{Parameter Error}

\begin{figure}[ht]
\centering
\includegraphics[width=\linewidth]{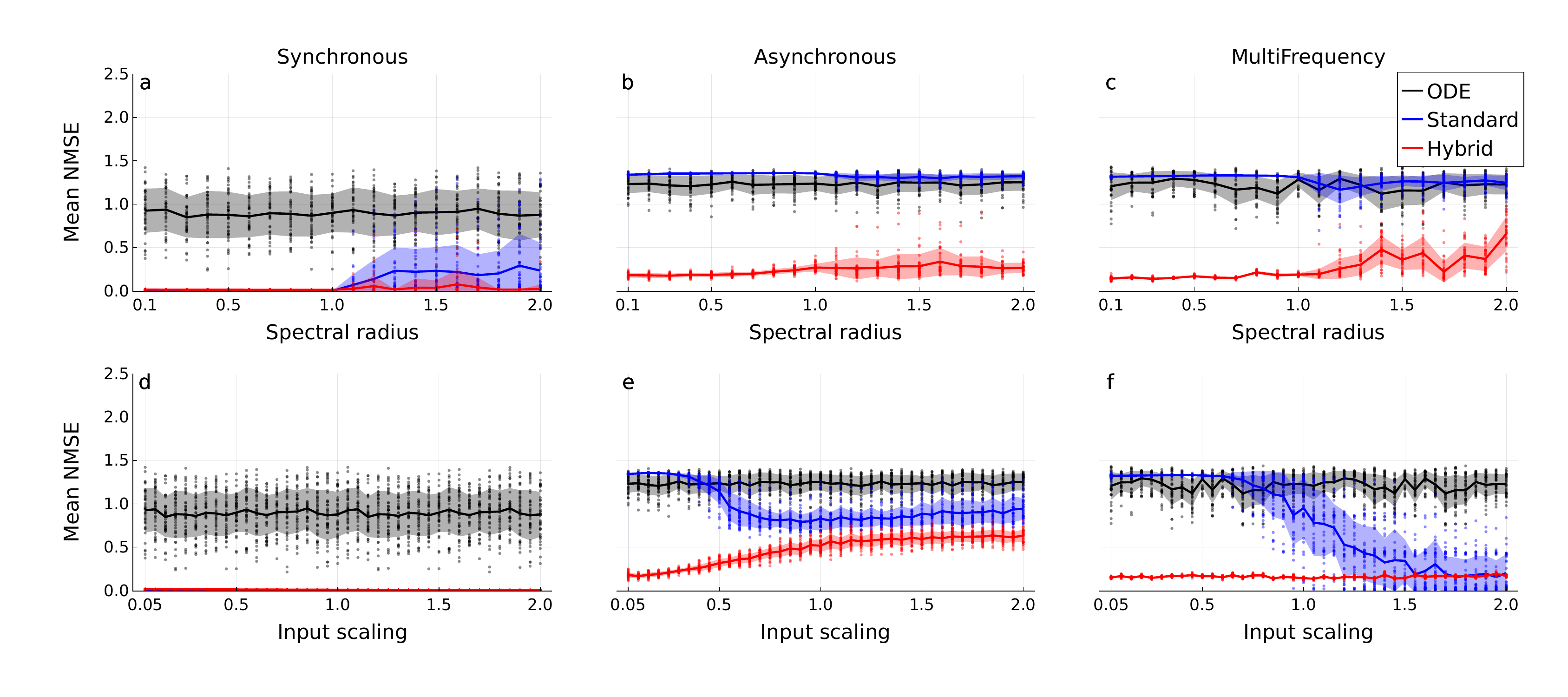}
\caption{Parameter-error task parameter sweeps evaluating the hybrid RC's prediction of NLON trajectories with parameter error in its expert ODE model. Mean NMSE in the prediction of the hybrid RC (\textcolor{red}{red}), standard RC (\textcolor{blue}{blue}), and the base ODE model (black) across the three dynamical regimes. Columns: Dynamical regime, synchronous (\textbf{a}, \textbf{d}), asynchronous (\textbf{b}, \textbf{d}), multi-frequency (\textbf{c}, \textbf{f}). Row: Parameter varied, spectral radius (\textbf{a}, \textbf{b}, \textbf{c}), input scaling (\textbf{d}, \textbf{e}, \textbf{f}). Individual dots are individual reservoir/ODE instantiations, each representing the mean NMSE across 60 forecasts, (20 for each realization of a ground truth regime). Solid lines are the mean of the mean NMSE across the reservoir/ODE instantiations. Shaded regions are one standard deviation across reservoir/ODE instantiations. The hybrid RC consistently outperforms the standard RC and the base ODE model. In the synchronous regime, a spectral radius above $1.0$ causes a degradation in standard RC performance. This is present for the hybrid RC but it recovers as spectral radius increases further. Increasing input scaling improves standard RC performance on both the asynchronous and multi-frequency regimes. At high input scaling on the multi-frequency regime, the standard RC matches the hybrid RC performance.}
\label{fig:task_1_sweeps}
\end{figure}

Our initial aim was to assess whether hybrid RCs are a viable architecture for surrogate modeling of NLONs by testing the long-term prediction quality across a range of parameters. We find that the performance of the hybrid RC is consistently better than the base ODE model or the standard RC. The hybrid RC achieves low error even when both of its constituent parts, the base ODE model and the standard RC, do not. This matches what was previously reported for the Lorenz system \cite{pathak2018hybrid}. Despite some variation due to parameter tuning, the hybrid RC often achieves near zero mean NMSE, particularly for the synchronous and multi-frequency regimes (Fig.~\ref{fig:task_1_sweeps} \textbf{a}, \textbf{d}, \textbf{c}, \textbf{f}). 

In the synchronous regime (\textbf{a}) there is an immediate departure from zero mean NMSE as the spectral radius crosses $1.0$ for the standard RC. Notably, the performance of the hybrid RC also suffers here, becoming more variable; however, perhaps surprisingly, it recovers as the spectral radius is increased further. For the asynchronous (\textbf{b}) and multi-frequency regimes (\textbf{c}), the standard RC and ODE models perform equally poorly, and again a slight increase in the error variance for a spectral radius greater than $1.0$ can be seen for both the standard and hybrid RC.

Changing the input scaling has little effect when predicting the synchronous (\textbf{d}) regime. However, for the asynchronous (\textbf{e}) and multi-frequency (\textbf{f}) regimes, increasing the input scaling improves the performance of the standard RC. Interestingly, the performance of the hybrid RC decreases with increasing input scaling for the asynchronous regime, but it still maintains an advantage over the standard RC throughout. In the multi-frequency regime, the performance of the hybrid RC is roughly constant, and the standard RC eventually matches it at an input scaling greater than ${1.7}$. The standard RC performance is generally more variable still, however some standard RCs achieve zero mean NMSE under this parameter setting. 
\subsection*{Residual Physics}
\begin{figure}[ht!]
\centering
\includegraphics[width=\linewidth]{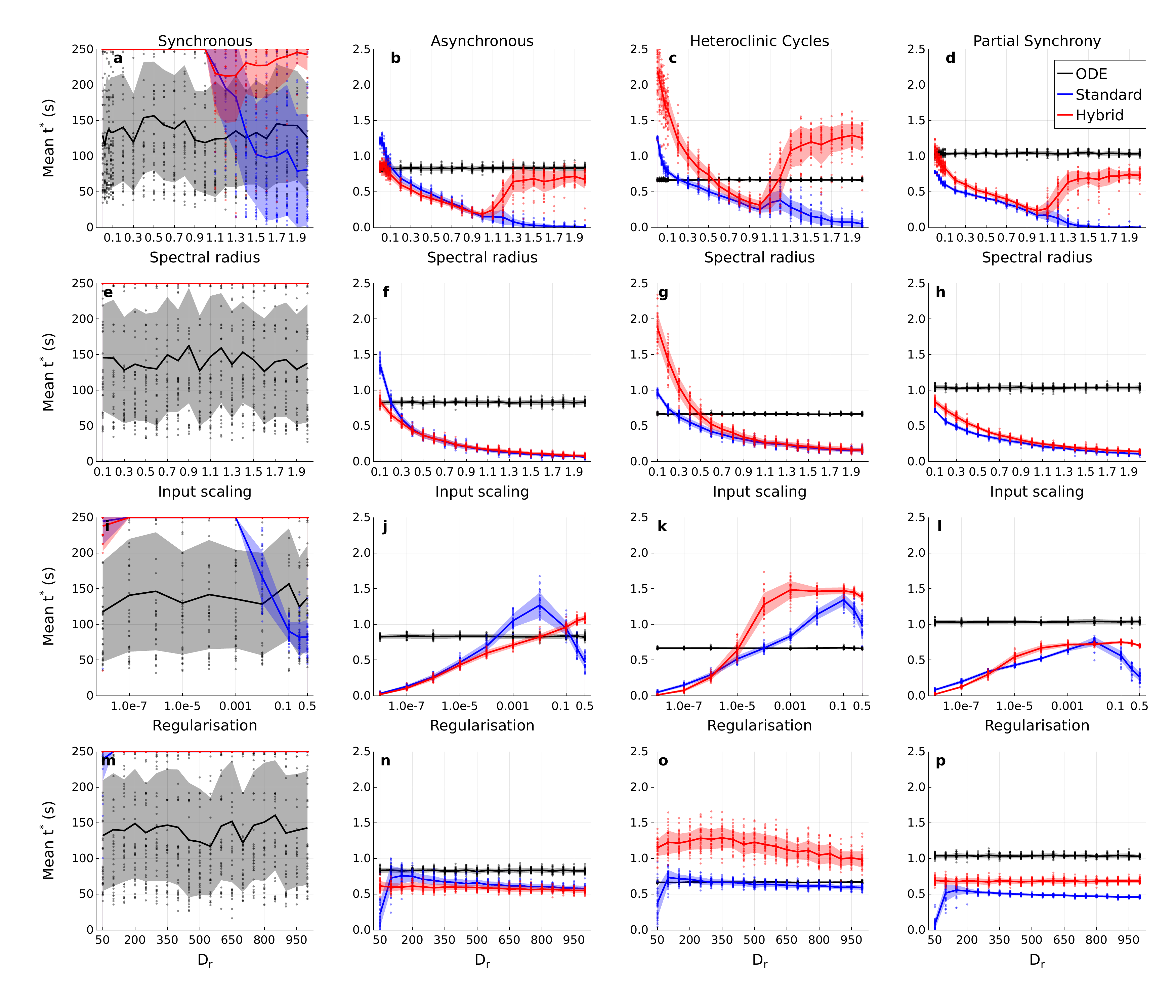}
\caption{Residual physics task parameter sweeps evaluating the short-term forecast performance of the hybrid RC with missing non-linearities in its expert ODE model. Mean valid time, $t^*$, achieved by the hybrid RC (\textcolor{red}{red}), standard RC (\textcolor{blue}{blue}), and the base ODE model (black) across the four dynamical regimes, and four different parameter sweeps. Column: Dynamical regime, synchronous (\textbf{a}, \textbf{e}, \textbf{i}, \textbf{m}), asynchronous (\textbf{b}, \textbf{f}, \textbf{j}, \textbf{n}), heteroclinic cycles~(\textbf{c}, \textbf{g}, \textbf{k}, \textbf{o}), partial synchrony~(\textbf{d}, \textbf{h}, \textbf{l}, \textbf{p}). Row: Parameter varied, spectral radius (\textbf{a}, \textbf{b}, \textbf{c}, \textbf{d}), input scaling (\textbf{e}, \textbf{f}, \textbf{g}, \textbf{h}), regularization (\textbf{i}, \textbf{j}, \textbf{k}, \textbf{l}), reservoir size, $D_r$, (\textbf{m}, \textbf{n}, \textbf{o}, \textbf{p}). Individual dots are individual reservoir/ODE instantiations, each representing the mean valid time across 20 forecasts. Solid lines are the mean of the mean valid time across the reservoir/ODE instantiations. Shaded regions are one standard deviation across reservoir/ODE instantiations. The hybrid RC generally outperforms the standard RC even on regimes out of its expert model's domain. The spectral radius significantly affects performance, with the standard RC showing degradation above a spectral radius of $1.0$. The hybrid RC behaves differently, showing recovery of performance when the spectral radius is above $1.0$. Input scaling primarily affects the asynchronous, \HC~and \SCPS~regimes, with optimum performance reached at minimal input scaling. Regularization strongly affects the performance of both the hybrid and standard RCs, with low regularization causing failure. Hybrid RCs have a broader range of viable regularization strength~on the~\HC~and~\SCPS~regimes. Reservoir size has little effect as long as a minimum of 100 nodes are available.}
\label{fig:valid_time_sweep}
\end{figure}
\subsubsection*{Parameter Sweep}
As in the parameter-error task we find that the hybrid RC generally performs better than the standard RC (Fig.~\ref{fig:valid_time_sweep}), although this is more variable than in the parameter-error task and the performance of the two is more similar. This is true even on the \textit{out of domain} heteroclinic cycles and partial synchrony regimes not accessible by the expert model alone. In the asynchronous case, the standard RC outperforms the hybrid RC, except for cases with spectral radii above 1.0. The base ODE model achieves higher valid times than the standard or hybrid RC across the partial synchrony regime sweeps (Fig.~\ref{fig:valid_time_sweep} \textbf{d}, \textbf{h}, \textbf{i}, \textbf{p}), and occasionally for the asynchronous (Fig.~\ref{fig:valid_time_sweep} \textbf{b}, \textbf{f}, \textbf{j}, \textbf{n}) and heteroclinic cycles regimes (Fig.~\ref{fig:valid_time_sweep} \textbf{c}, \textbf{g}, \textbf{k}, \textbf{o}). It is clear however, that the long term prediction of the base ODE model is not accurate as it fails to capture the qualitative character of these regimes (Supplementary Figures ~S11, ~S12, ~S13).
	
The spectral radius again has a strong effect on the performance of the standard and hybrid RC; furthermore, the effect is different for the two (Fig.~\ref{fig:valid_time_sweep} \textbf{a},\textbf{ b}, \textbf{c}, \textbf{d}). Most notable is the divergence in behavior at and above a spectral radius of $1.0$. There is a sharp decline in performance for the standard RC on the synchronous regime similar to that observed in the parameter error task (Fig.~\ref{fig:task_1_sweeps} \textbf{a}). The standard RC achieves the maximum $250~\text{s}$ valid time with a spectral radius below $1.0$ but this falls to ~$75~\text{s}$ for a spectral radius of $2.0$. As in the parameter error task, the hybrid RC begins to fail at a spectral radius of $1.0$ but then recovers as the spectral radius is increased further  (\textbf{a}). On all the other regimes, (\textbf{b}, \textbf{c}, \textbf{d}), the hybrid RC shows the same recovery, this time after a steady decrease in valid time as the spectral radius increases to $1.0$, followed by an increase back to near its maximum valid time.~On these regimes optimum performance is reached for the standard and hybrid RC's with a spectral radius of $0.0$. The input scaling parameter (Fig.~\ref{fig:valid_time_sweep} \textbf{e}, \textbf{f}, \textbf{g}, \textbf{h}) only affects the more complex dynamical regimes (\textbf{f} - asynchronous, \textbf{g} - \HC, \textbf{h} - \SCPS), with maximum valid time for the standard and hybrid RC at minimum input scaling.

The regularization parameter (Fig.~\ref{fig:valid_time_sweep}: \textbf{i}, \textbf{j}, \textbf{k}, \textbf{l}) has a strong effect on the valid time of the standard and hybrid RC. High regularization in the synchronous regime greatly reduces the performance of the standard RC. In the asynchronous, \HC,~and \SCPS~dynamical regimes, sufficiently low regularization appears to cause failure of the standard and hybrid RC. In all cases, the optimal regularization strength range is much broader for the hybrid RC than the standard RC.

Beyond requiring at least $100$ nodes, the reservoir size (Fig \ref{fig:valid_time_sweep} \textbf{m}, \textbf{n}, \textbf{o}, \textbf{p}) has little effect on the valid time on any of the regimes. The hybrid RC benefits slightly from a larger reservoir ($350$ nodes) when predicting the \HC~regime but it still outperforms the standard RC across the full range of reservoir sizes explored.

\subsubsection*{Grid Search}

\begin{figure}[ht!]
\centering
\includegraphics[width=\linewidth]{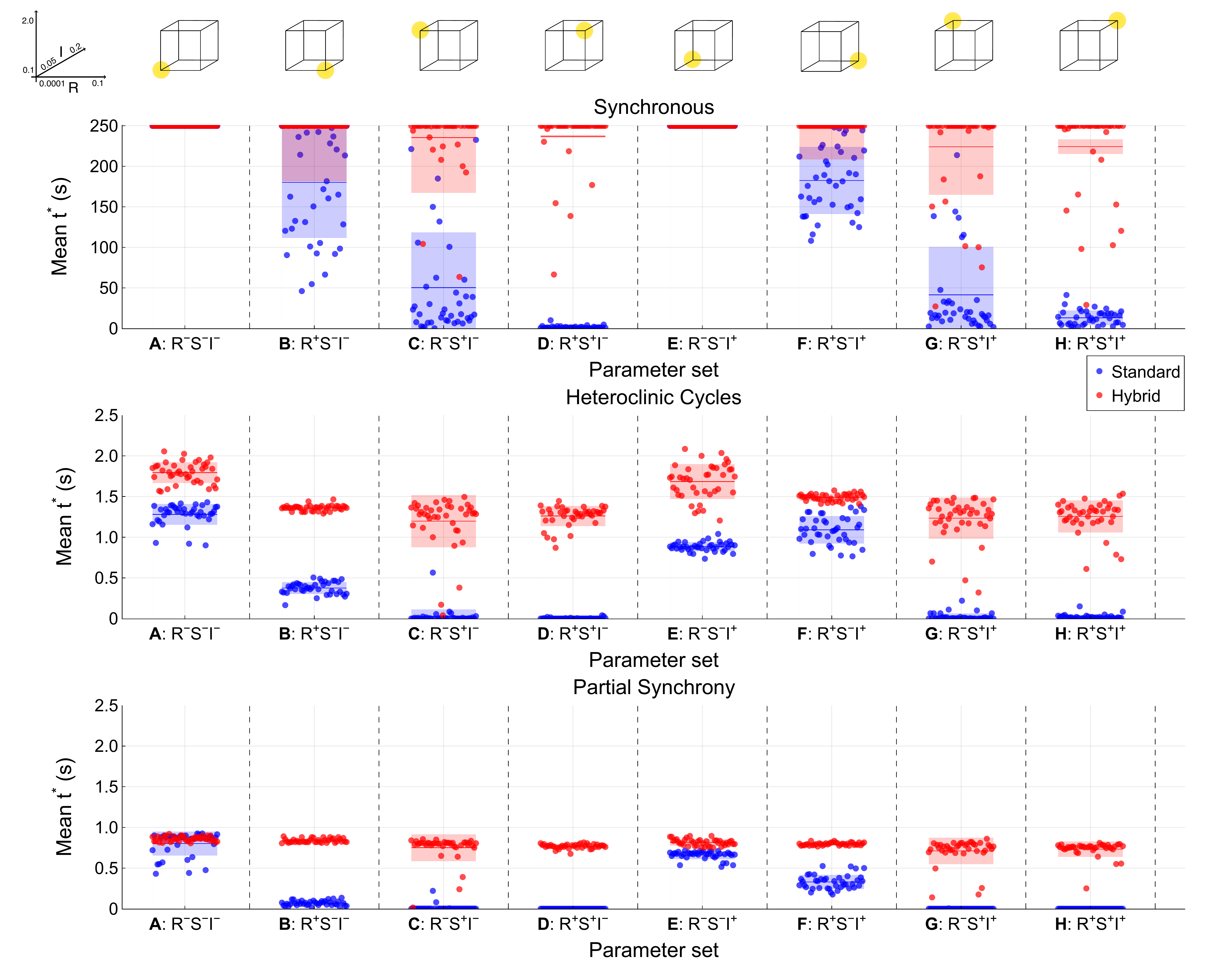}
\caption{Mean valid time, $t^*$, achieved by individual hybrid (\textcolor{red}{red}) and standard (\textcolor{blue}{blue}) RCs across 20 test forecasts in the grid search. Columns: Parameter sets corresponding to each corner of the grid search cube are labeled \textbf{A} to \textbf{H}, (Fig.~\ref{fig:grid_search_cube}). Rows: Dynamical regimes. Top row: synchronous, middle row: heteroclinic cycles, bottom row: partial synchrony. Horizontal lines: the mean of the mean valid time across reservoirs. Shaded regions: one standard deviation across reservoirs. On the synchronous regime, Hybrid RCs achieve near perfect performance for all parameter sets. The standard RC also performs well on the synchronous regime, however high regularization and spectral radius significantly reduce performance. On the \HC~regime, hybrid RCs consistently outperform standard RCs. High spectral radius again reduces the performance of the standard RC. Low regularization increases the variance in the hybrid RC performance. Maximum valid time on the \HC~regime is $45.8\%$ higher for the hybrid RC than the standard RC. On the \SCPS~regime, both models perform poorly, however the hybrid RC is more consistent across parameter sets. High spectral radius severely reduces the standard RC performance again. The hybrid RC does not show an improvement in maximum valid time over the standard RC on the partial synchrony regime.}
\label{fig:grid_search_results}
\end{figure}

In the final part of this study we investigate how easy it is to optimize hybrid RCs. We evaluated the performance of hybrid and standard RCs using the valid time metric in a grid search of the hyper parameters that were shown to most strongly affect performance (Fig.~\ref{fig:valid_time_sweep}). We ran tests on three of the dynamical regimes (synchronous, \HC~and \SCPS), evaluating the hybrid RC and standard RC with parameter settings defined by the corners of a cube in parameter space, labeled \textbf{A} to \textbf{G} (Fig.~\ref{fig:grid_search_cube}).

As identified in the parameter sweep tests, the synchronous regime is the easiest to predict (Fig.~\ref{fig:grid_search_results}, top row). The hybrid RC achieves near perfect valid times ($250~\text{s}$) across all parameter settings in the grid search. A high spectral radius (\textbf{C}, \textbf{D}, \textbf{G}, \textbf{H}) slightly reduces the performance of some hybrid reservoirs, but the majority still reach $250~\text{s}$ valid time. In contrast, the standard RC is strongly affected by a high spectral radius as observed earlier in the parameter sweeps, with many reservoirs achieving less than $50~\text{s}$ valid time (\textbf{C}, \textbf{D}, \textbf{G}, \textbf{H}). High regularization strength (\textbf{B}, \textbf{F}) also reduces the standard RC performance, but not as significantly as high spectral radius. Input scaling has little effect on the performance here as the chosen settings are only slightly above and below the optimum value found during the parameter sweeps for both the hybrid RC and standard RC (0.1). Optimal performance on the synchronous regime is reached with a low setting of all three hyper-parameters, with all standard and hybrid RCs reaching $250~\text{s}$ valid time (\textbf{A}, \textbf{E}).

In the heteroclinic cycles regime (Fig.~\ref{fig:grid_search_results}, middle row), the hybrid RCs consistently outperform the standard RCs. The performance is poor however, with valid times on the order of $1.0$ to $2.0~\text{s}$ ($10$ to $20$ steps). The individual hybrid RCs reach higher valid times than the standard RCs in all cases except parameter setting C, with low regularization strength, high spectral radius, and low input scaling. A high spectral radius (\textbf{C}, \textbf{D}, \textbf{G}, \textbf{H}) significantly reduces the performance of the standard RC. The hybrid RC is less affected by this, aside from a noticeable increase in variance when regularization is high (\textbf{B} to \textbf{D}, and \textbf{F} to \textbf{H}). The maximum valid time reached by the hybrid RC is $2.085~\text{s}$, and for the standard RC it is $1.435~\text{s}$. This is a $45.8\%$ improvement for the `optimal' hybrid RC over the `optimal' standard RC.

The performance of the standard and hybrid RC is least favorable on the \SCPS~regime (Fig.~\ref{fig:grid_search_results}, bottom row), achieving valid times only up to $1.0~\text{s}$ ($10$ steps). There is still a clear performance difference between the two RC models however. In their best parameter setting (\textbf{A} - low spectral radius, input scaling, and regularization), the performance of the two models is similar, but the hybrid RC has less variance across reservoir instantiations. A high spectral radius again results in poor performance for the standard RC, with many standard RCs achieving $0~\text{s}$ valid time (\textbf{C}, \textbf{D}, \textbf{G}, \textbf{H}). Comparing the statistics of the valid time across parameter settings, the mean hybrid RC performance is fairly stable and the performance of individual reservoirs tightly grouped, whereas the mean performance of the standard RC is at or near $0~\text{s}$ for all but parameter settings \textbf{A}, \textbf{E}, and \textbf{F}. The maximum valid time reached by the best hybrid RC is $0.920~\text{s}$, and for the best standard RC it is $0.925~\text{s}$. This is not an improvement over the standard RC, however the hybrid RC approaches this far more consistently than the standard RC.

\section*{Discussion}
While the performance advantage of hybrid RC over standard RC and the base ODE is known from previous work \cite{pathak2018hybrid}, we have extended this in two ways. First, we reproduced this result but applied to the important example of an oscillating system. Second, we introduced a new task, the residual physics task, to mimic situations where the interaction between oscillators is not fully known. This was a more stringent test than the parameter error task, however we still observe higher performance from the hybrid RC than the standard RC, although, using the valid time metric, the base ODE does better in some cases.~Unlike the clear advantage seen in ~\cite{pathak2018hybrid}, the benefit of the hybrid approach is thus more nuanced here. As soon as the ground truth regime becomes complex, the performance of the reservoirs is substantially degraded, dropping from $250~\text{s}$ valid time, to around $1~\text{s}$, which is low even compared to the valid times achieved on the chaotic Lorenz system ($\sim10~\text{s}$ \cite{pathak2018hybrid}), although these predictions may still be useful.

Notably, although both the standard and hybrid RC had low valid times on the~\HC~and~\SCPS~regimes, the hybrid RC consistently outperformed the standard RC. This suggests that the hybrid RC captures some aspects of the residual physics, with, for example, a relative maximum performance gain of $45.8\%$ on the \HC~regime. Although the valid times are low for these regimes, they may be sufficient for control applications with short time horizon requirements.

There are clear limitations of the valid time metric for any application requiring long term, qualitative prediction accuracy, such as attractor reconstruction tasks. This is evident in the qualitative differences between the base ODE and the hybrid RC's trajectory predictions for asynchronous,~\HC, and~\SCPS~regimes (Supplementary Fig.~S11, S12, and S13), which do not align with their relative performance with valid time. The ODE quickly falls into a synchronous regime, predicting synchronized, slow trajectories at odds with the ground truth for the~asynchronous, heteroclinic cycles, or partial synchrony regimes. The addition of the reservoir in the hybrid RC changes this, giving oscillations that much better capture the gross features of the ground truth behavior. In the heteroclinic cycles regime, the hybrid RC more accurately recreates the underlying frequency of the main cluster, and produces oscillators that briefly leave this attracting group. In the partial synchrony regime, the hybrid RC again more closely reproduces the underlying oscillation frequency of the distribution. However it struggles to capture some complex features of these regimes, such as oscillator departure timing in the heteroclinic cycles regime, and the distributional behavior of the partial synchrony regime. When they fail, both the standard and hybrid RC can produce noisy, high-frequency oscillations, or steady state trajectories (Supplementary Fig.~S16). As such, the robustness of the hybrid RC may be useful in safety critical applications, where oscillation decay or growth could be catastrophic.

Over-fitting is a risk for any data-based modeling but is a particular hazard when using models for control. The standard RC requires careful tuning of the regularization strength to avoid this whereas the hybrid RC does not, even when using an expert model missing substantial components. This is particularly true for the \HC~and \SCPS~regimes, where the performance of the hybrid RC is good across a broader range of regularization strengths, above and below the optimum value required for the standard RC. The output matrix of the hybrid RC balances attention between the reservoir activity and the expert model's predictions. By providing extra states for weight assignment containing some correct features of the ground truth, the hybrid RC reduces training data influence and over-fitting risk.~This is a delicate, task dependent, balance, as evidenced by Figure~\ref{fig:valid_time_sweep} (\textbf{b}, \textbf{f}, \textbf{j}, \textbf{n}), where the standard RC outperforms the hybrid RC, but the hybrid RC performance increases with increasing regularization strength. Inspection of the predicted trajectories for this regime (Supplementary Fig.~S11), indicates the hybrid RC may be biased too far towards the synchronous behavior of its expert model with the baseline regularization setting.

We have shown that the scaling of the reservoir weights has a strong effect on prediction quality for the standard RC; the hybrid RC is also affected, but in a different and less pronounced way. The spectral radius, controlling the scale of the internal connectivity matrix, is particularly influential. This is a well-documented phenomenon for a standard RC \cite{venayagamoorthy2009effects,jiang2019model}. The scaling parameters are tied to prediction quality through their impact on the RC's Lyapunov exponents, and conditional Lyapunov exponents (CLEs). These are features of dynamical systems that determine the structure of phase space, and any constituent attractors. 
Primarily, they are critical for the satisfaction of the echo state property, a fundamental property for the function of an RC that imbues a reservoir with a fading memory and capacity for generalized signal-induced synchronization \cite{maass2002real,pecora2015synchronization}. Heuristically, it has been suggested that a spectral radius below one is required to have the echo state property, however this is neither necessary nor sufficient \cite{yildiz2012re}. The spectral radius is however correlated with the magnitude of the RC's CLEs \cite{hart2024attractor}, and we observe failure of the standard RC for spectral radii greater than one in accordance with this. 

The hybrid RC does not respond in the same way. Lyapunov exponents and CLEs are also key for successful attractor reconstruction \cite{pathak2017using,hart2024attractor}. Although our performance metric is reporting short-term prediction accuracy and not the long-term ergodic property accuracy, the \textit{attractor climate}, we suggest that the hybrid RC's divergent response to increasing spectral radius could be a result of a different correspondence between spectral radius and the CLEs leading to successful attractor reconstruction where the standard RC has failed.

For optimum performance, a spectral radius of zero, corresponding to no memory of past states and no coupling between reservoir nodes, is most effective. This is unsurprising as the ground truth system is first order, and all state information is available to the reservoirs; all the information required for the next step can be obtained from the current step alone.

The scaling parameters have also been implicated in the connection between RC and delay embedding \cite{duan2023embedding,gauthier2021next}. The high performance of both standard and hybrid RCs at low spectral radius suggests that they do not require a long memory for the tasks considered here. To investigate whether the reservoirs might benefit from richer representations of fewer past states instead of long memory, we tested reservoirs that completed multiple internal updates per step. These ``multi-step'' reservoirs are similar to the ``drifting-state'' reservoirs in~~\cite{sakemi2020model} except only the final internal state is used for output computation, after multiple non-linear updates have been applied to form a more complex representation. We found no consistent effect on performance across various dynamical system forecasting tasks (Supplementary Fig.~S17).

The behavior of the standard and hybrid RC in response to parameter changes is similar across the three non-synchronous regimes. A further test conducted with a slower version of the asynchronous regime resulted in near-zero valid times for both models across all parameters (Supplementary Fig.~S18). We believe this was due to the slow, highly correlated, dynamics (Supplementary Fig.~S15) creating a training span in a distinct region of phase space from the test spans (Supplementary Fig.~S19).

The improved performance of the hybrid RC over the standard RC comes at the cost of increased computational complexity. Further work must evaluate, on a case-by-case basis, whether the performance improvement justifies the cost, and whether the computational cost alone rules out chosen control applications. Extending the hybrid RC architecture to include recent developments in reservoir computing \cite{jaurigue2021reservoir,koster2023data,ma2023novel} may help to reduce this cost.

\section*{Data Availability}
The data that support the findings of this study are available from the corresponding author upon reasonable request.

\section*{Code Availability}
The code used to generate and process the results of this study are available from the corresponding author upon reasonable request.

\section*{Acknowledgments}
AS is funded by a UK Research and Innovation grant (EP/S022937/1). This work was carried out using the computational facilities of the Advanced Computing Research Centre, University of Bristol - http://www.bris.ac.uk/acrc/.

\section*{Author contributions statement}
AS, CH, DB and MH: conceptualization, methodology, writing - review \& editing. CH, DB and MH: project management. AS: investigation, formal analysis, visualization, writing - original draft preparation.

\section*{Competing interests} 
The authors declare no competing interests.
\clearpage
\section*{Supplementary Information}
\renewcommand{\thesection}{S\arabic{section}}
\renewcommand{\thetable}{S\arabic{table}}
\renewcommand{\thefigure}{S\arabic{figure}}
\renewcommand{\figurename}{Supplementary Figure}
\setcounter{figure}{0}
\large{\textbf{Modeling Nonlinear Oscillator Networks Using Physics-Informed Hybrid Reservoir Computing}}

\noindent Andrew Shannon, Conor Houghton, David Barton, Martin Homer
\begin{figure}[ht!]
\centering
\includegraphics[width=\linewidth]{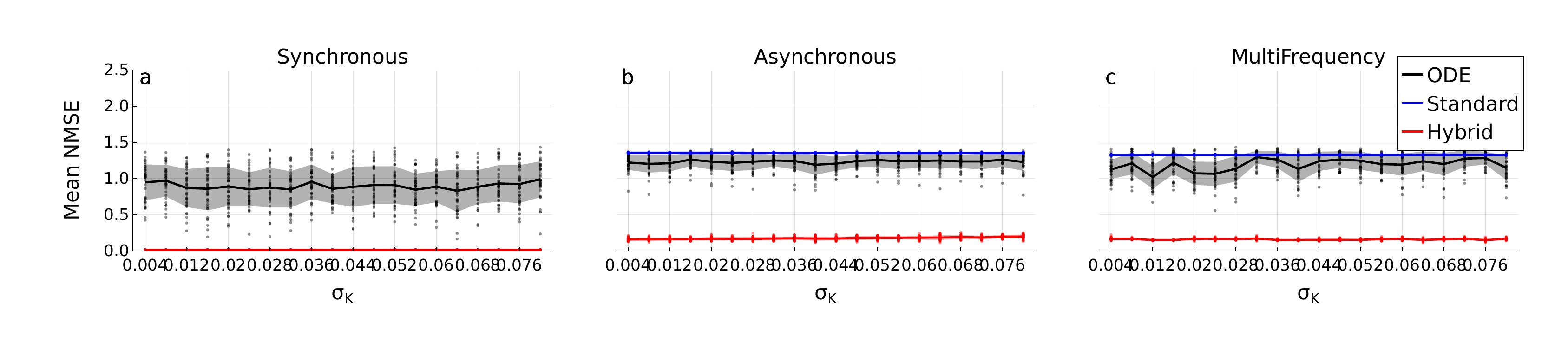}
\caption{Parameter error task parameter sweeps evaluating the hybrid RC's prediction of NLON trajectories with parameter error in its expert model as the standard deviation of the coupling strength error, $\sigma_K$, is varied. Mean NMSE in the prediction of the hybrid RC (\textcolor{red}{red}), standard RC (\textcolor{blue}{blue}), and the base ODE model (black) across the three dynamical regimes. Column - dynamical regime: Synchronous (\textbf{a}), Asynchronous (\textbf{b}), Multi-Frequency (\textbf{c}). Individual dots are individual reservoir/ODE instantiations (40), each representing the mean NMSE across 60 forecasts, (20 for each realization of a ground truth regime). Solid lines are the mean NMSE across the reservoir/ODE instantiations. Shaded regions are one standard deviation across reservoir/ODE instantiations.}
\label{supp_fig:K_sweep}
\end{figure}

\begin{figure}[ht!]
\centering
\includegraphics[width=\linewidth]{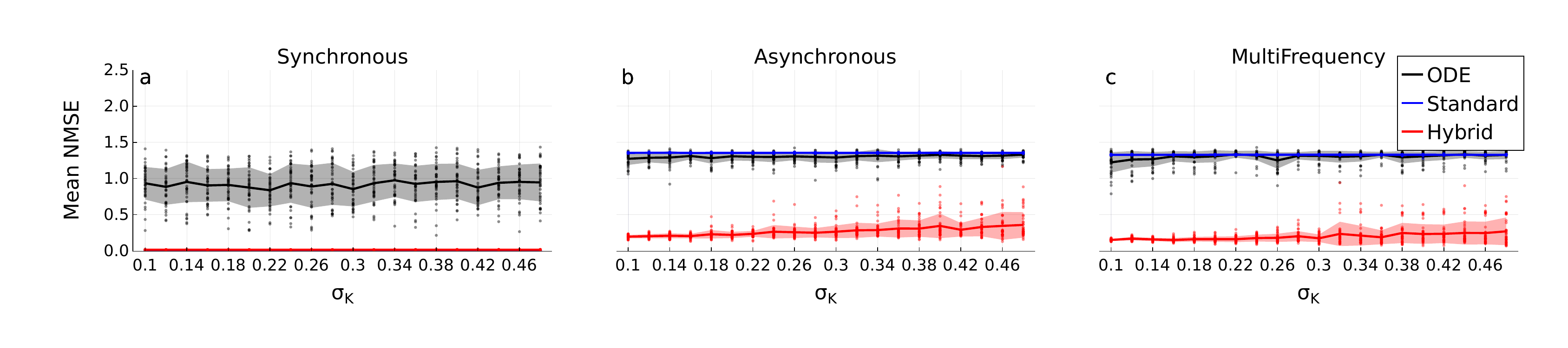}
\caption{Parameter error task parameter sweeps evaluating the hybrid RC's prediction of NLON trajectories with parameter error in its expert model as the standard deviation of the coupling strength error, $\sigma_K$,is varied across high values. Mean NMSE in the prediction of the hybrid RC (\textcolor{red}{red}), standard RC (\textcolor{blue}{blue}), and the base ODE model (black) across the three dynamical regimes. Column - dynamical regime: Synchronous (\textbf{a}), Asynchronous (\textbf{b}), Multi-Frequency (\textbf{c}). Individual dots are individual reservoir/ODE instantiations (40), each representing the mean NMSE across 60 forecasts, (20 for each realization of a ground truth regime). Solid lines are the mean NMSE across the reservoir/ODE instantiations. Shaded regions are one standard deviation across reservoir/ODE instantiations.}
\label{supp_fig:K_large_sweep}
\end{figure}

\begin{figure}[ht!]
\centering
\includegraphics[width=\linewidth]{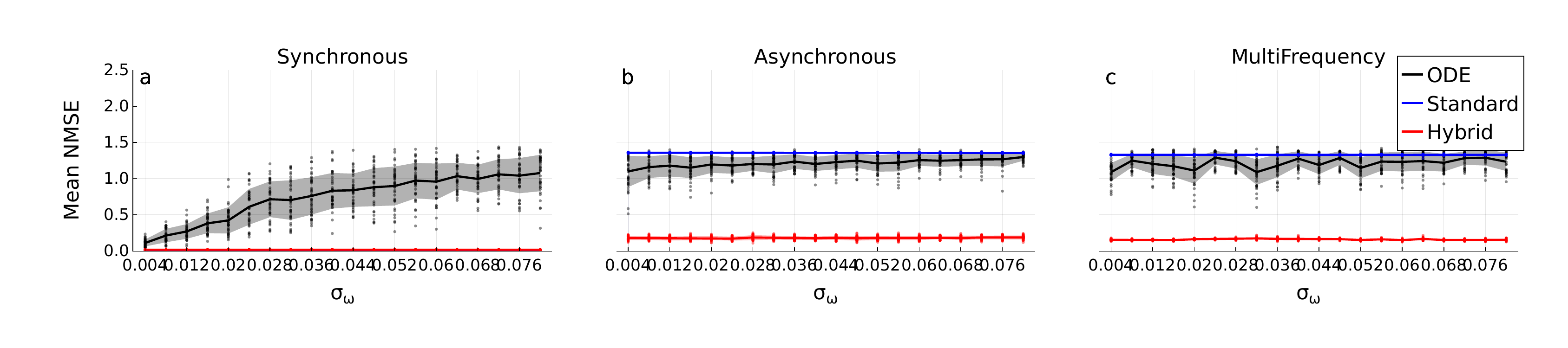}
\caption{Parameter error task parameter sweeps evaluating the hybrid RC's prediction of NLON trajectories with parameter error in its expert model as the standard deviation of the natural frequency error, $\sigma_{\omega}$, is varied. Mean NMSE in the prediction of the hybrid RC (\textcolor{red}{red}), standard RC (\textcolor{blue}{blue}), and the base ODE model (black) across the three dynamical regimes. Column - dynamical regime: Synchronous (\textbf{a}), Asynchronous (\textbf{b}), Multi-Frequency (\textbf{c}). Individual dots are individual reservoir/ODE instantiations (40), each representing the mean NMSE across 60 forecasts, (20 for each realization of a ground truth regime). Solid lines are the mean NMSE across the reservoir/ODE instantiations. Shaded regions are one standard deviation across reservoir/ODE instantiations.}
\label{supp_fig:Om_sweep}
\end{figure}

\begin{figure}[ht!]
\centering
\includegraphics[width=\linewidth]{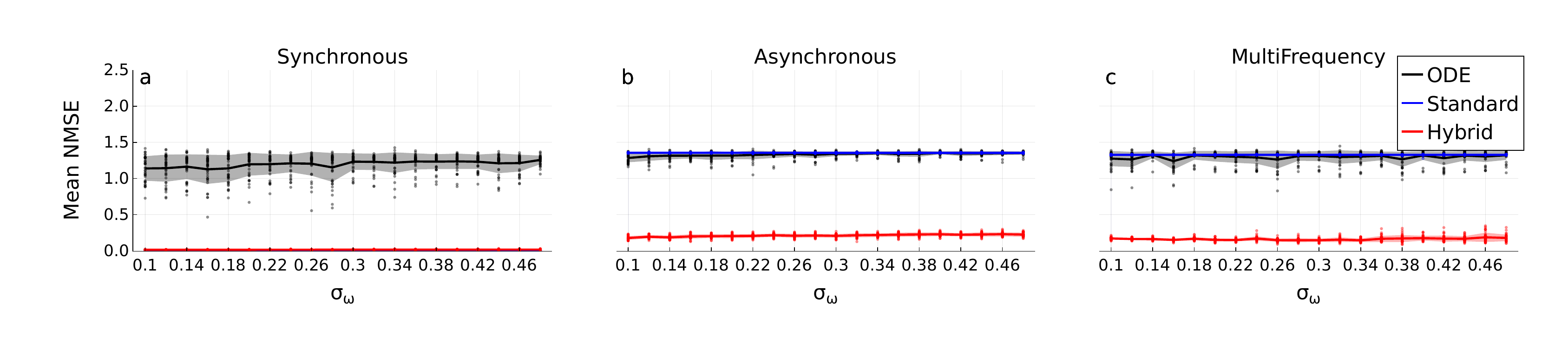}
\caption{Parameter error task parameter sweeps evaluating the hybrid RC's prediction of NLON trajectories with parameter error in its expert model as the standard deviation of the natural frequency error, $\sigma_{\omega}$, is varied across high values. Mean NMSE in the prediction of the hybrid RC (\textcolor{red}{red}), standard RC (\textcolor{blue}{blue}), and the base ODE model (black) across the three dynamical regimes. Column - dynamical regime: Synchronous (\textbf{a}), Asynchronous (\textbf{b}), Multi-Frequency (\textbf{c}). Individual dots are individual reservoir/ODE instantiations (40), each representing the mean NMSE across 60 forecasts, (20 for each realization of a ground truth regime). Solid lines are the mean NMSE across the reservoir/ODE instantiations. Shaded regions are one standard deviation across reservoir/ODE instantiations.}
\label{supp_fig:Om_large_sweep}
\end{figure}

\begin{figure}[ht!]
\centering
\includegraphics[width=\linewidth]{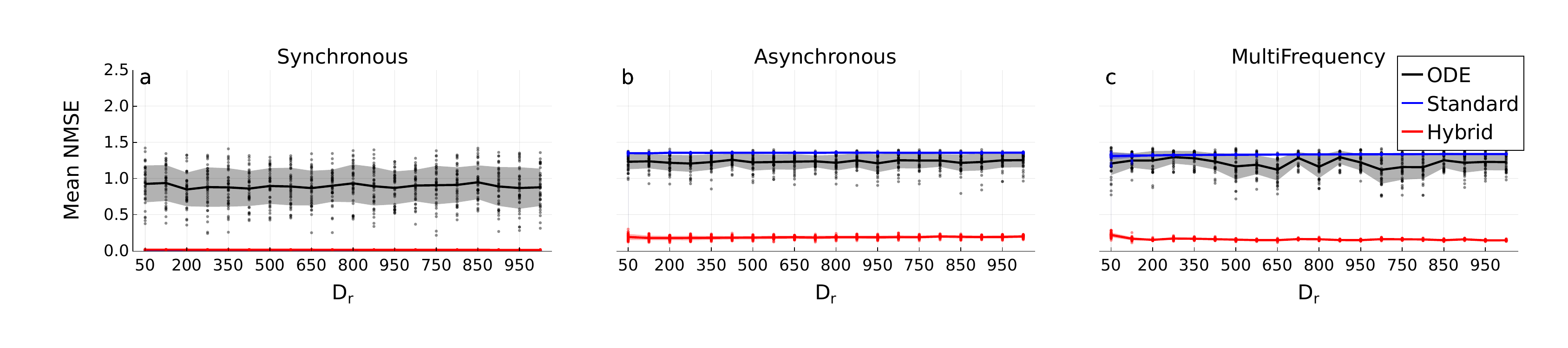}
\caption{Parameter error task parameter sweeps evaluating the hybrid RC's prediction of NLON trajectories with parameter error in its expert model as the reservoir size is varied. Mean NMSE in the prediction of the hybrid RC (\textcolor{red}{red}), standard RC (\textcolor{blue}{blue}), and the base ODE model (black) across the three dynamical regimes. Column - dynamical regime: Synchronous (\textbf{a}), Asynchronous (\textbf{b}), Multi-Frequency (\textbf{c}). Individual dots are individual reservoir/ODE instantiations (40), each representing the mean NMSE across 60 forecasts, (20 for each realization of a ground truth regime). Solid lines are the mean NMSE across the reservoir/ODE instantiations. Shaded regions are one standard deviation across reservoir/ODE instantiations.}
\label{supp_fig:ResSize_sweep}
\end{figure}

\begin{figure}[ht!]
\centering
\includegraphics[width=\linewidth]{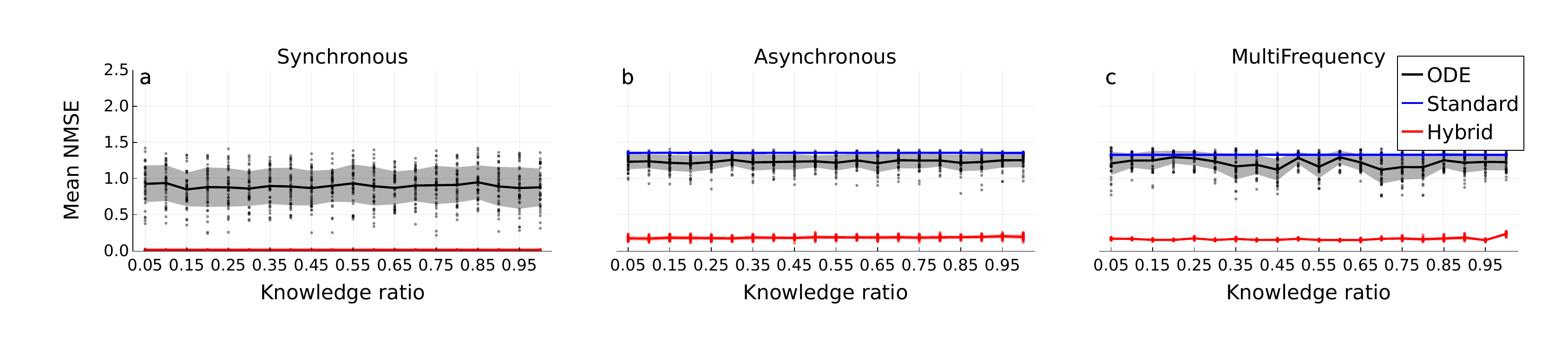}
\caption{Parameter error task parameter sweeps evaluating the hybrid RC's prediction of NLON trajectories with parameter error in its expert model as the knowledge ratio is varied. Mean NMSE in the prediction of the hybrid RC (\textcolor{red}{red}), standard RC (\textcolor{blue}{blue}), and the base ODE model (black) across the three dynamical regimes. Column - dynamical regime: Synchronous (\textbf{a}), Asynchronous (\textbf{b}), Multi-Frequency (\textbf{c}). Individual dots are individual reservoir/ODE instantiations (40), each representing the mean NMSE across 60 forecasts, (20 for each realization of a ground truth regime). Solid lines are the mean NMSE across the reservoir/ODE instantiations. Shaded regions are one standard deviation across reservoir/ODE instantiations.}
\label{supp_fig:KR_sweep}
\end{figure}

\begin{figure}[ht!]
\centering
\includegraphics[width=\linewidth]{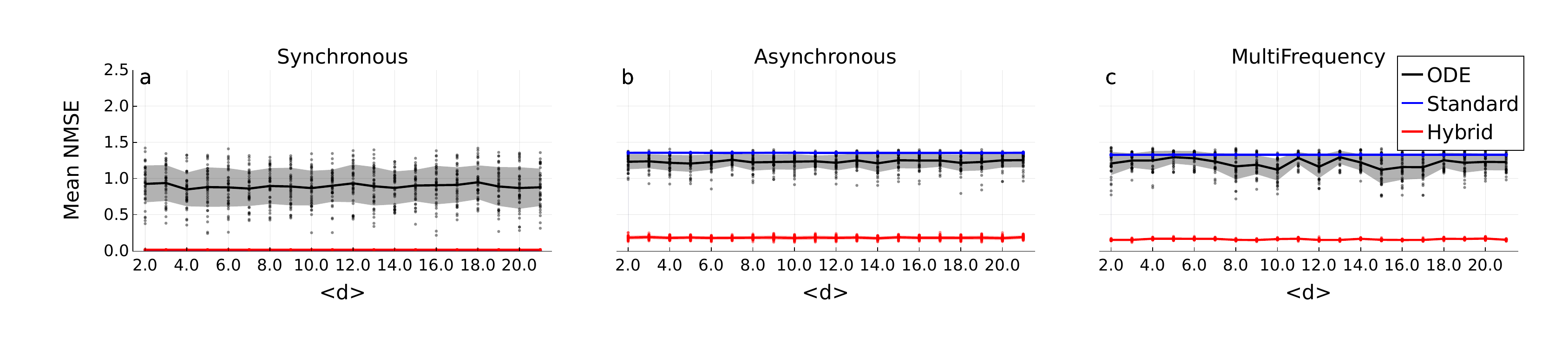}
\caption{Parameter error task parameter sweeps evaluating the hybrid RC's prediction of NLON trajectories with parameter error in its expert model as the mean degree is varied. Mean NMSE in the prediction of the hybrid RC (\textcolor{red}{red}), standard RC (\textcolor{blue}{blue}), and the base ODE model (black) across the three dynamical regimes. Column - dynamical regime: Synchronous (\textbf{a}), Asynchronous (\textbf{b}), Multi-Frequency (\textbf{c}). Individual dots are individual reservoir/ODE instantiations (40), each representing the mean NMSE across 60 forecasts, (20 for each realization of a ground truth regime). Solid lines are the mean NMSE across the reservoir/ODE instantiations. Shaded regions are one standard deviation across reservoir/ODE instantiations.}
\label{supp_fig:MeanDegree_sweep}
\end{figure}

\begin{figure}[ht!]
\centering
\includegraphics[width=\linewidth]{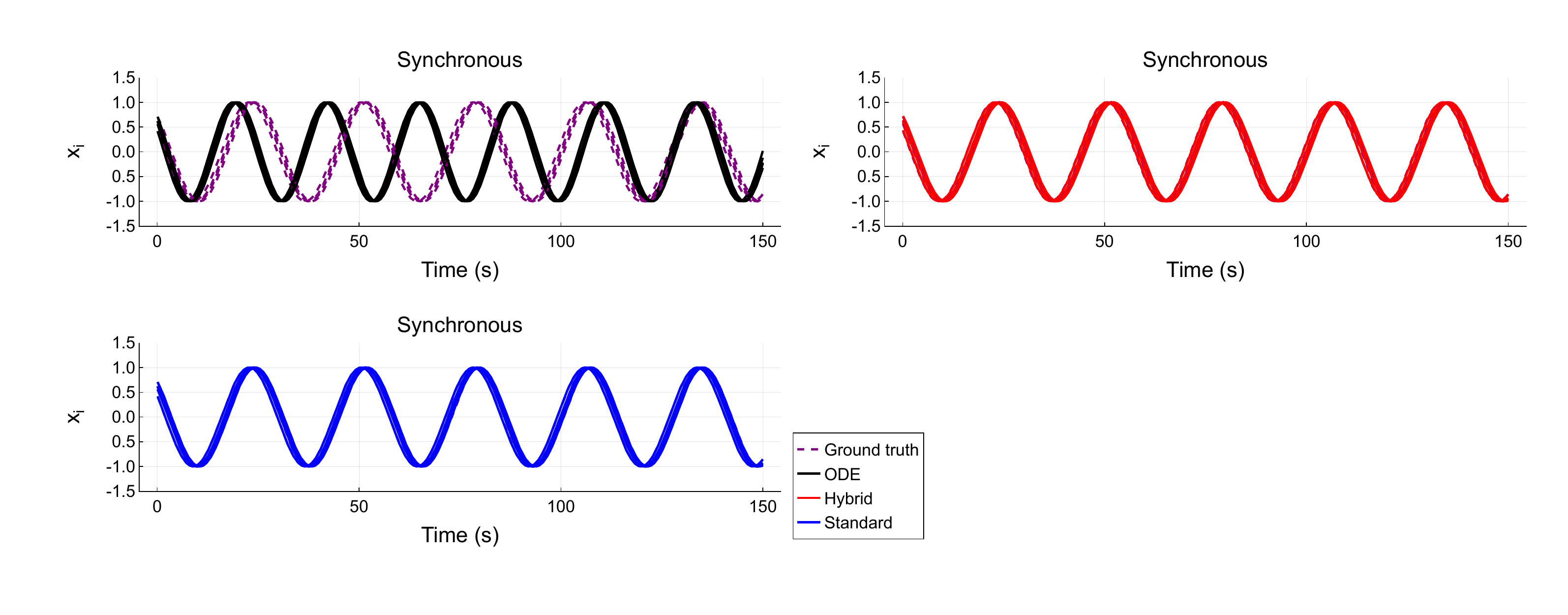}
\caption{Example synchronous regime phase-component trajectories forecast by each model compared to the ground truth from the parameter error task's standard Kuramoto model. Base ODE (black), standard RC (\textcolor{blue}{blue}), hybrid RC (\textcolor{red}{red}), ground truth (\textcolor{purple}{dashed purple}). Both the standard and hybrid RCs accurately forecast the synchronous dynamics. The ODE does produce synchronous dynamics, but the error in its natural frequencies causes the forecast to drift. Parameters from the tenth index of the second $\sigma_K$ sweep corresponding to $\sigma_K=0.28$, all other parameters at baseline.}
\label{supp_fig:PE_sync_traj}
\end{figure}

\begin{figure}[ht!]
\centering
\includegraphics[width=\linewidth]{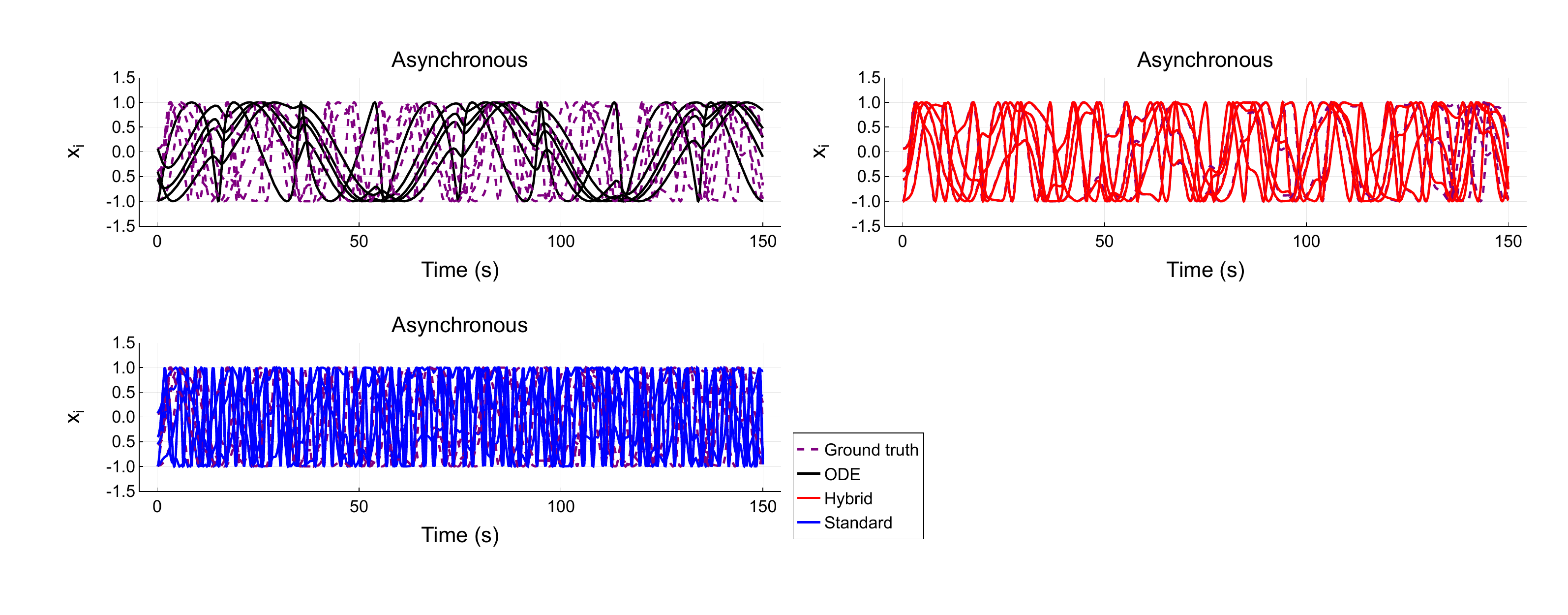}
\caption{Example asynchronous regime phase-component trajectories forecast by each model compared to the ground truth from the parameter error task's standard Kuramoto model. Base ODE (black), standard RC (\textcolor{blue}{blue}), hybrid RC (\textcolor{red}{red}), ground truth (\textcolor{purple}{dashed purple}). The error in both the coupling strength and natural frequencies of the ODE model is enough to stop it accurately forecasting the complex asynchronous dynamics. The hybrid RC does exceptionally well for this particular instantiation and parameter setting. In contrast, the standard RC does not. Parameters from the tenth index of the second $\sigma_K$ sweep corresponding to $\sigma_K=0.28$, all other parameters at baseline.}
\label{supp_fig:PE_async_traj}
\end{figure}

\begin{figure}[ht!]
\centering
\includegraphics[width=\linewidth]{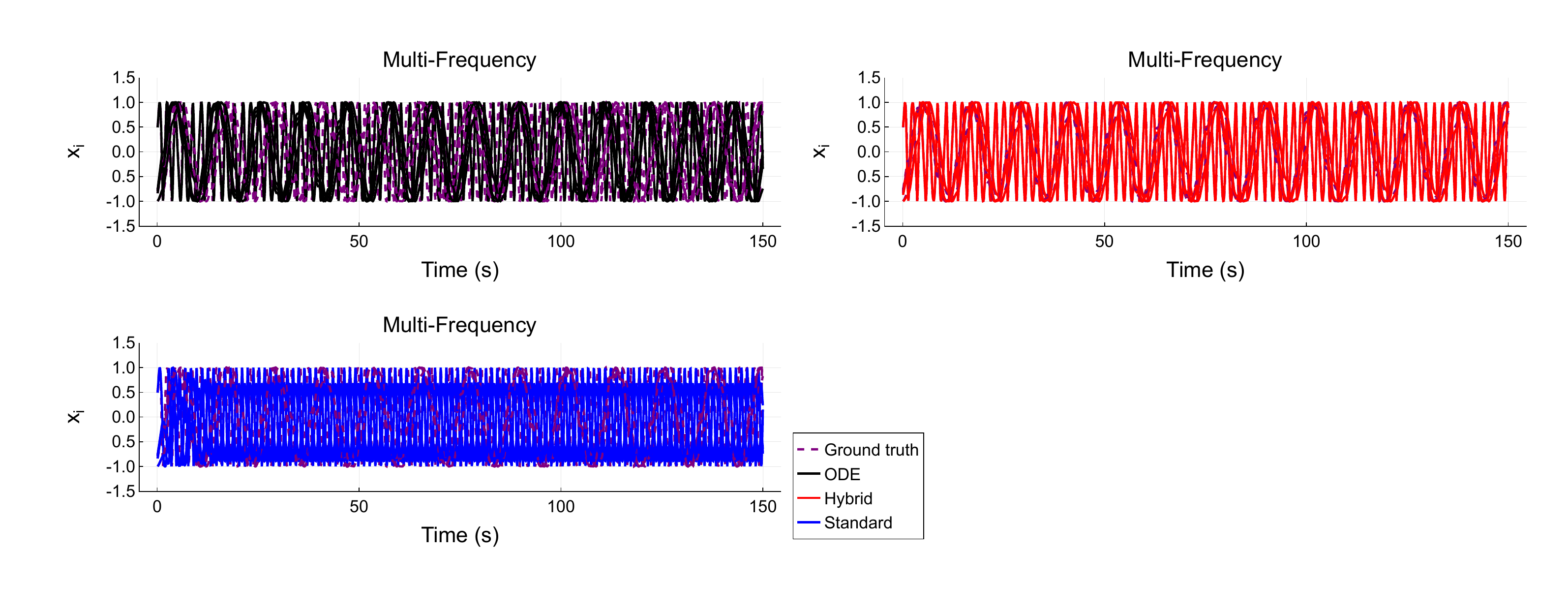}
\caption{Example multi-frequency regime phase-component trajectories forecast by each model compared to the ground truth from the parameter error task's standard Kuramoto model. Base ODE (black), standard RC (\textcolor{blue}{blue}), hybrid RC (\textcolor{red}{red}), ground truth (\textcolor{purple}{dashed purple}). As for the synchronous regime, the ODE model correctly captures the synchronous low frequency cluster's and phase-locked high frequency oscillator's behavior but drifts away due to the natural frequency error. The hybrid RC nearly perfectly forecasts the dynamics. The standard RC struggles with predicting the high frequency oscillator, in contrast to its excellent performance on the synchronous regime; the strong time-scale separation may be the cause of this. Parameters from the tenth index of the second $\sigma_K$ sweep corresponding to $\sigma_K=0.28$, all other parameters at baseline.}
\label{supp_fig:PE_MF_traj}
\end{figure}


\begin{figure}[ht!]
\centering
\includegraphics[width=\linewidth]{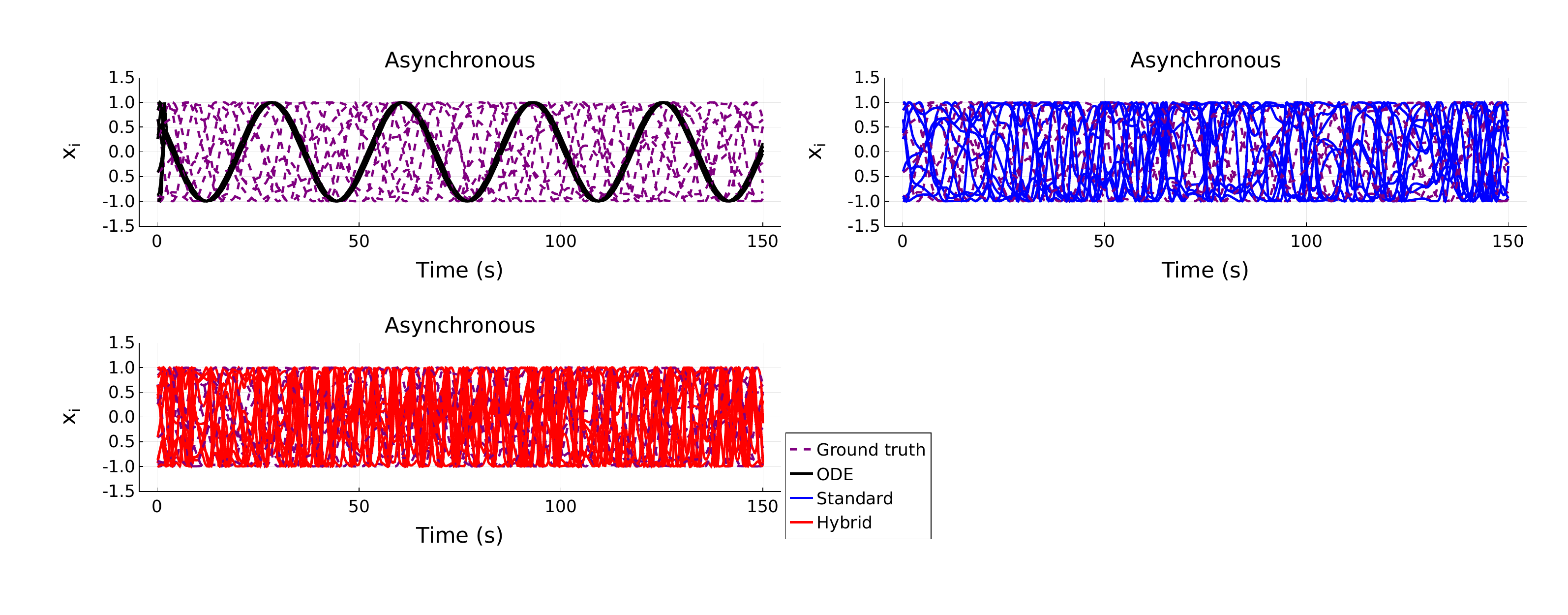}
\caption{Example asynchronous regime phase-component trajectories forecast by each model compared to the ground truth from the residual physics task's bi-harmonic Kuramoto model. Base ODE (black), standard RC (\textcolor{blue}{blue}), hybrid RC (\textcolor{red}{red}), ground truth (\textcolor{purple}{dashed purple}). The base ODE model forecast achieves a higher valid time than the hybrid and standard RC but is clearly failing to capture the main feature of the dynamics once this time has elapsed as it synchronizes. The reservoirs can produce more realistic oscillations, with the standard reservoir better capturing the qualitative dynamics for this parameter set. Parameters from the third index of the spectral radius sweep corresponding to a spectral radius of $0.3$, all other parameters at baseline.}
\label{supp_fig:asyncfast_traj}
\end{figure}

\begin{figure}[ht!]
\centering
\includegraphics[width=\linewidth]{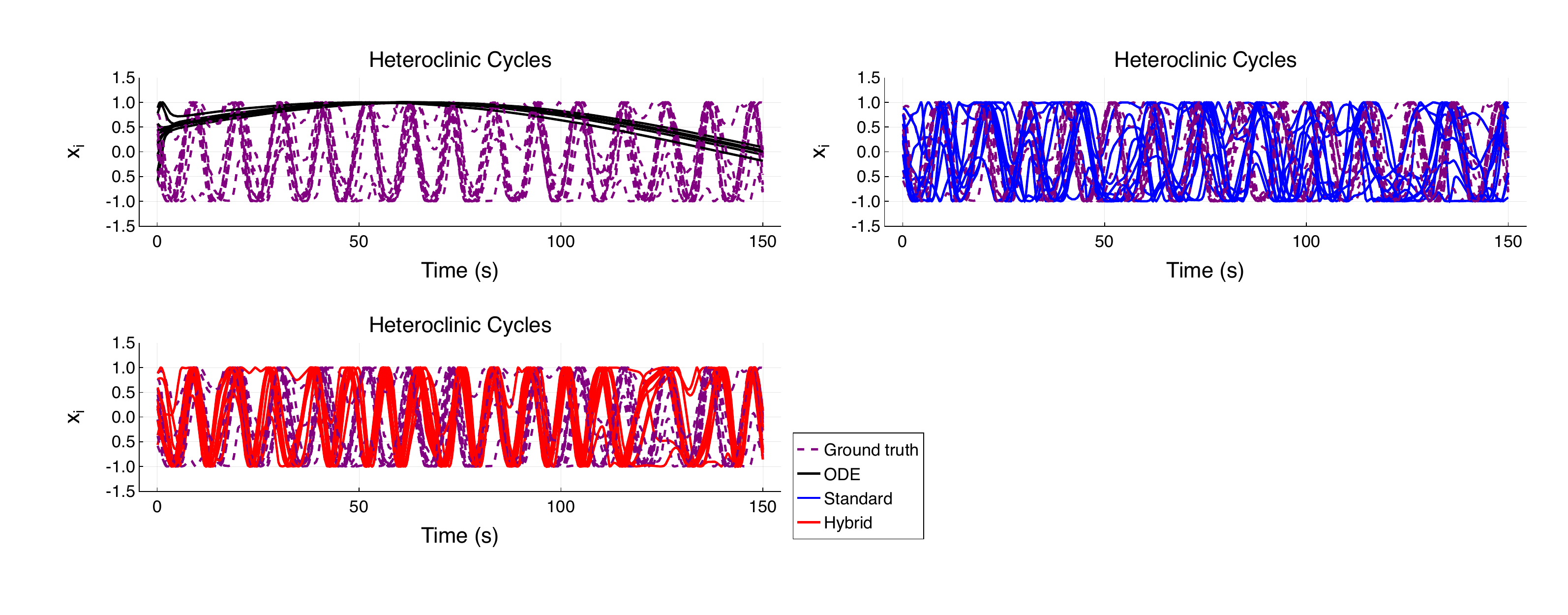}
\caption{Example heteroclinic cycles regime phase-component trajectories forecast by each model compared to the ground truth from the residual physics task's bi-harmonic Kuramoto model. Base ODE (black), standard RC (\textcolor{blue}{blue}), hybrid RC (\textcolor{red}{red}), ground truth (\textcolor{purple}{dashed purple}). The base ODE model predicts slow synchronous trajectories, whilst the standard and hybrid RC better recreate the gross features of the dynamics. The hybrid RC in particular seems to best capture the underlying oscillation frequency and also has some oscillators departing from and returning to the main attracting cluster as is required for this regime. Parameters from the first index of the input scaling sweep corresponding to an input scaling of $0.05$, all other parameters at baseline.}
\label{supp_fig:HC_traj}
\end{figure}

\begin{figure}[ht!]
\centering
\includegraphics[width=\linewidth]{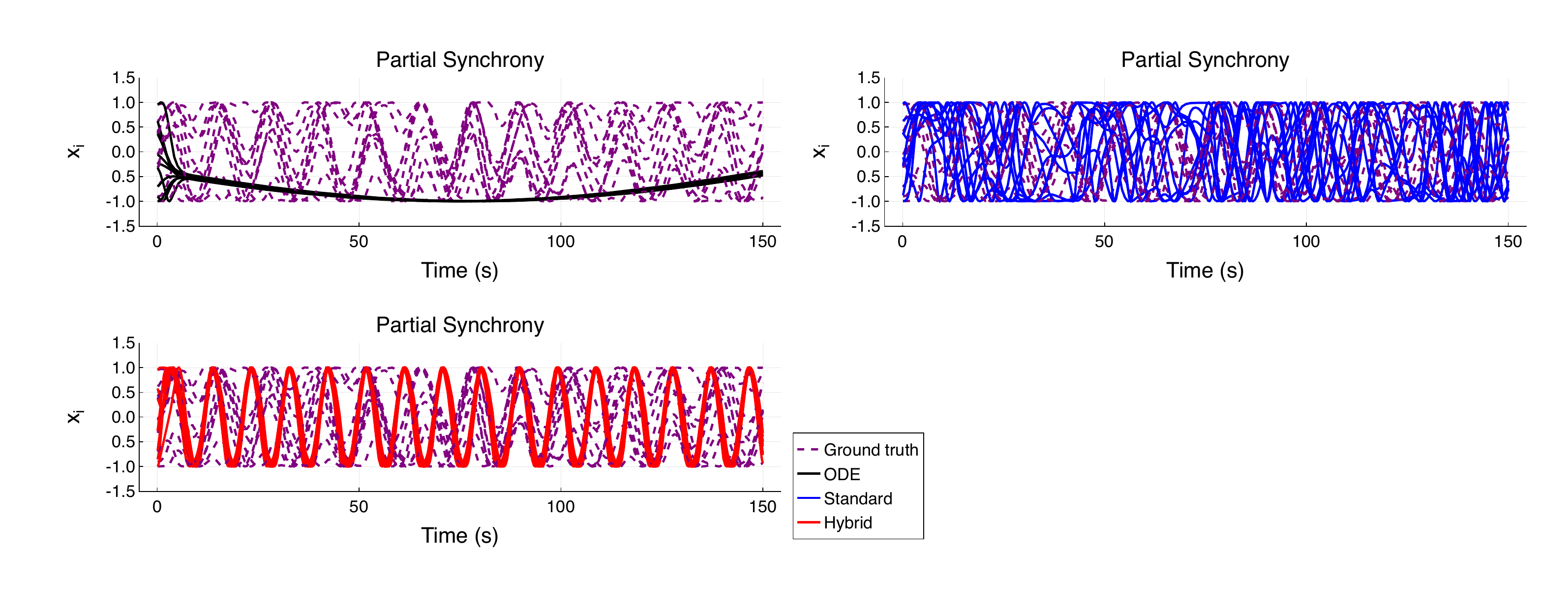}
\caption{Example partial synchrony regime phase-component trajectories forecast by each model compared to the ground truth from the residual physics task's bi-harmonic Kuramoto model. Base ODE (black), standard RC (\textcolor{blue}{blue}), hybrid RC (\textcolor{red}{red}), ground truth (\textcolor{purple}{dashed purple}). The base ODE model fails to capture the partial synchrony regime, predicting slow, synchronous dynamics. In contrast the hybrid RC successfully captures the underlying oscillation frequency. The hybrid RC forecasts overly synchronous trajectories however, failing to capture the partially synchronous, distributional behavior characteristic of this regime. Parameters from the first index of the input scaling sweep corresponding to an input scaling of $0.05$, all other parameters at baseline.}
\label{supp_fig:SCPS_traj}
\end{figure}

\begin{figure}[ht!]
\centering
\includegraphics[width=\linewidth]{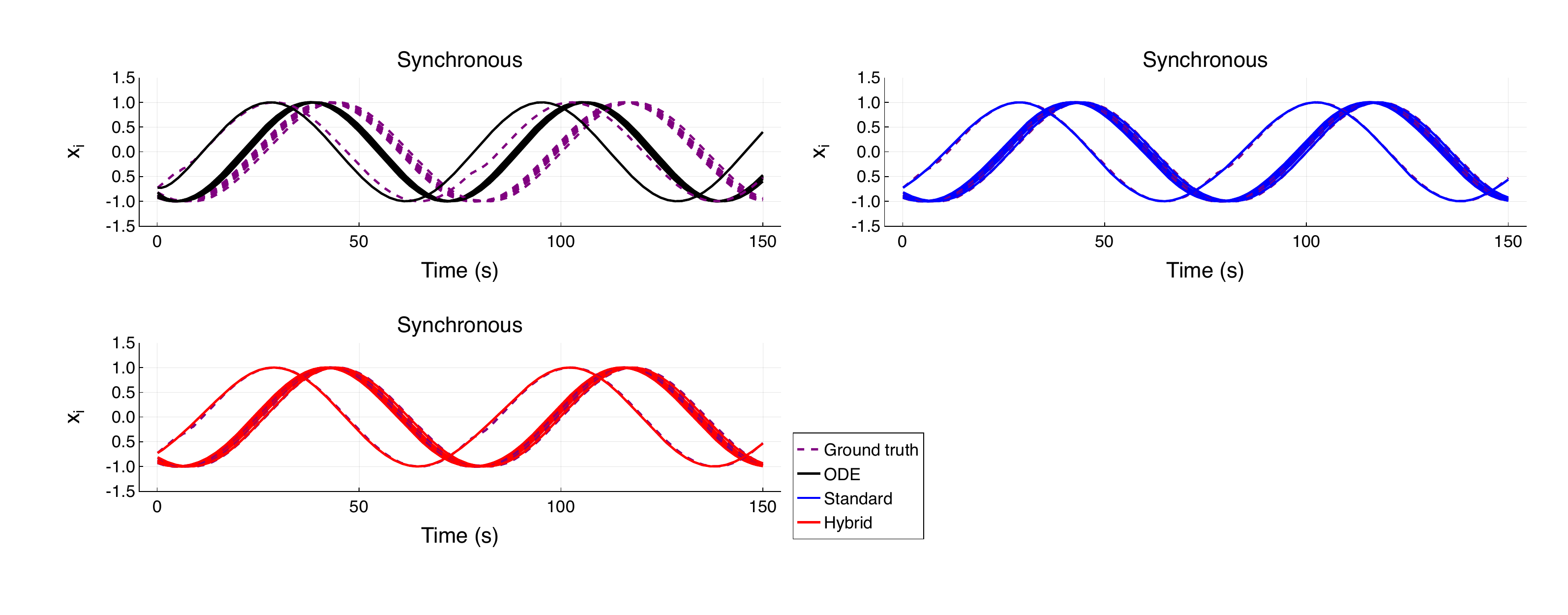}
\caption{Example synchronous regime phase-component trajectories forecast by each model compared to the ground truth from the residual physics task's bi-harmonic Kuramoto model. Base ODE (black), standard RC (\textcolor{blue}{blue}), hybrid RC (\textcolor{red}{red}), ground truth (\textcolor{purple}{dashed purple}). All three models do well on the synchronous regime, with the standard and hybrid RC often reaching the maximum $250$ second valid time. The base ODE however drifts due to parameter error. Parameters from the first index of the input scaling sweep corresponding to an input scaling of $0.05$, all other parameters at baseline.}
\label{supp_fig:sync_traj}
\end{figure}

\begin{figure}[ht!]
\centering
\includegraphics[width=\linewidth]{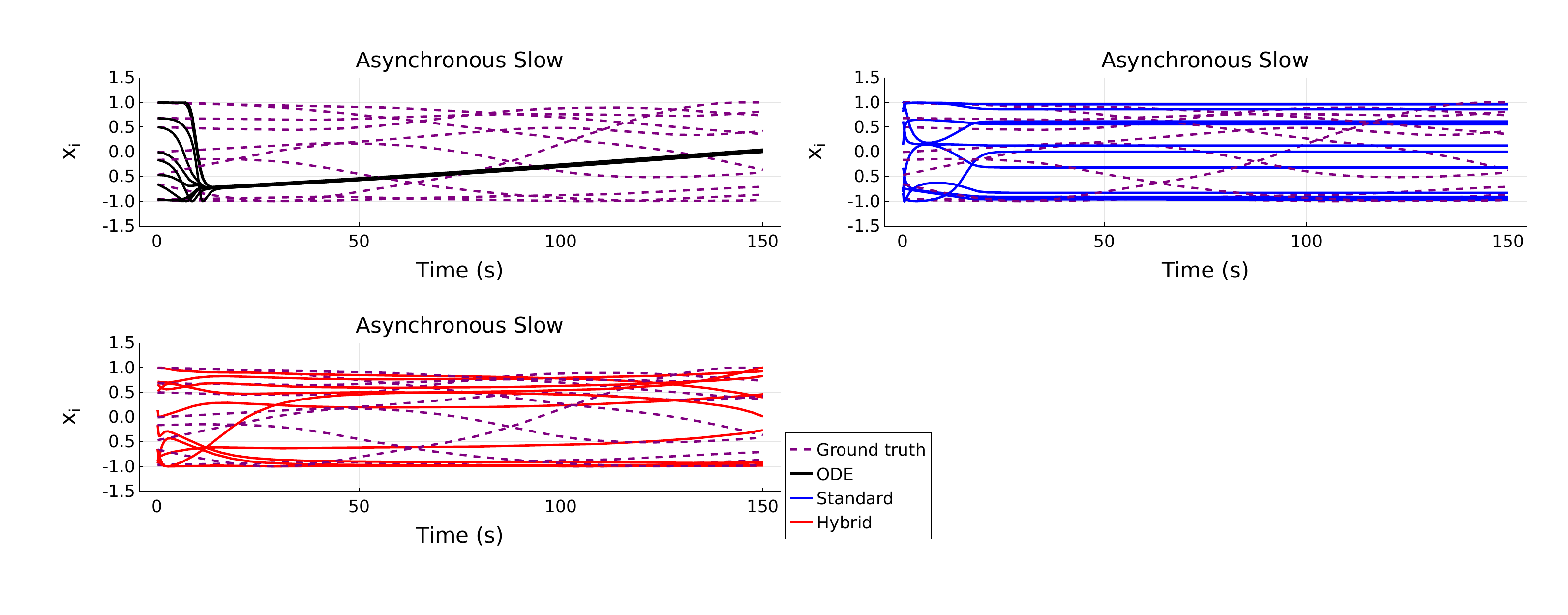}
\caption{Example slow asynchronous regime phase-component trajectories forecast by each model compared to the ground truth from the residual physics task's bi-harmonic Kuramoto model. Base ODE (black), standard RC (\textcolor{blue}{blue}), hybrid RC (\textcolor{red}{red}), ground truth (\textcolor{purple}{dashed purple}). The slow, synchronous base ODE model forecast achieves a higher valid time than the hybrid and standard RC but is clearly failing to capture the main feature of the dynamics once this time has elapsed. The standard RC also fails, but through what appears to be falling into a fixed point solution. Parameters from the first index of the input scaling sweep corresponding to an input scaling of $0.05$, all other parameters at baseline.}
\label{supp_fig:asyncslow_traj}
\end{figure}

\begin{figure}[ht!]
\centering
\includegraphics[width=\linewidth]{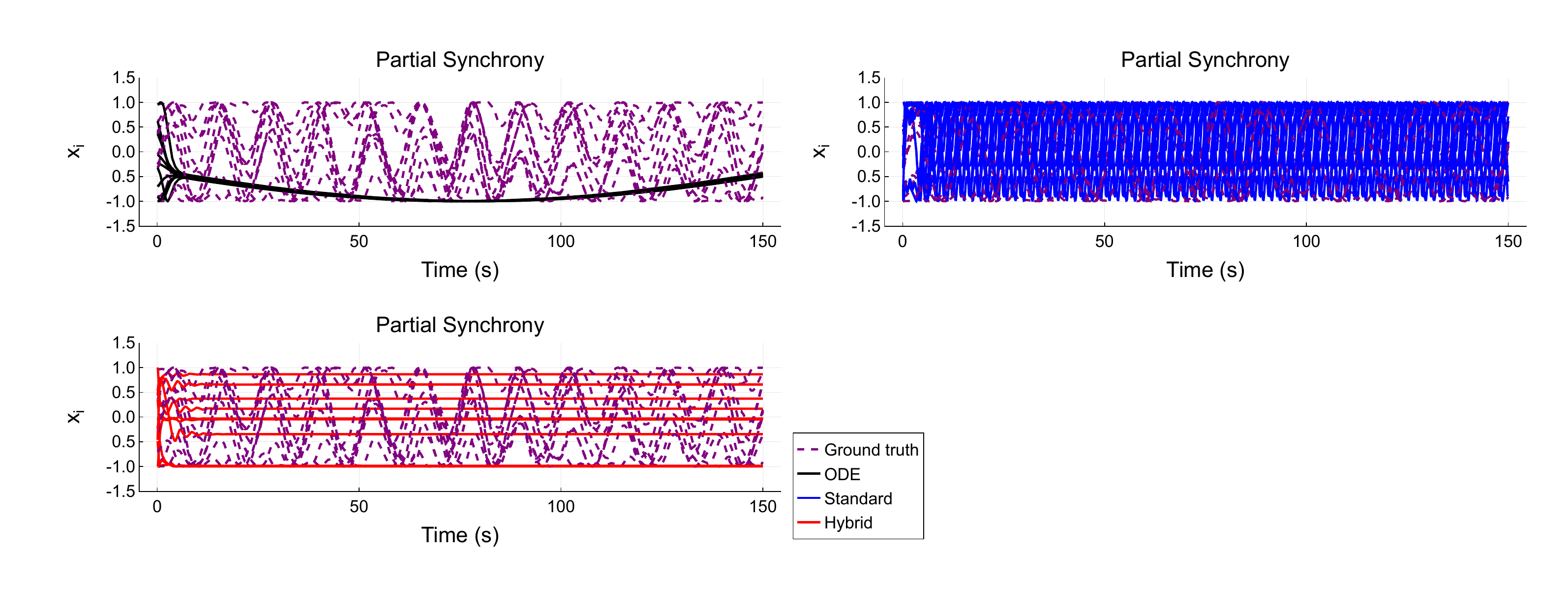}
\caption{Example failed partial synchrony regime phase-component trajectories forecast by each model compared to the ground truth from the residual physics task's bi-harmonic Kuramoto model. Base ODE (black), standard RC (\textcolor{blue}{blue}), hybrid RC (\textcolor{red}{red}), ground truth (\textcolor{purple}{dashed purple}). In this case (parameters at baseline with input scaling set to 1.9) the hybrid and standard RC's fail by producing steady state and high frequency trajectories respectively.}
\label{supp_fig:failures}
\end{figure}


\begin{figure}[ht!]
\centering
\includegraphics[width=0.7\linewidth]{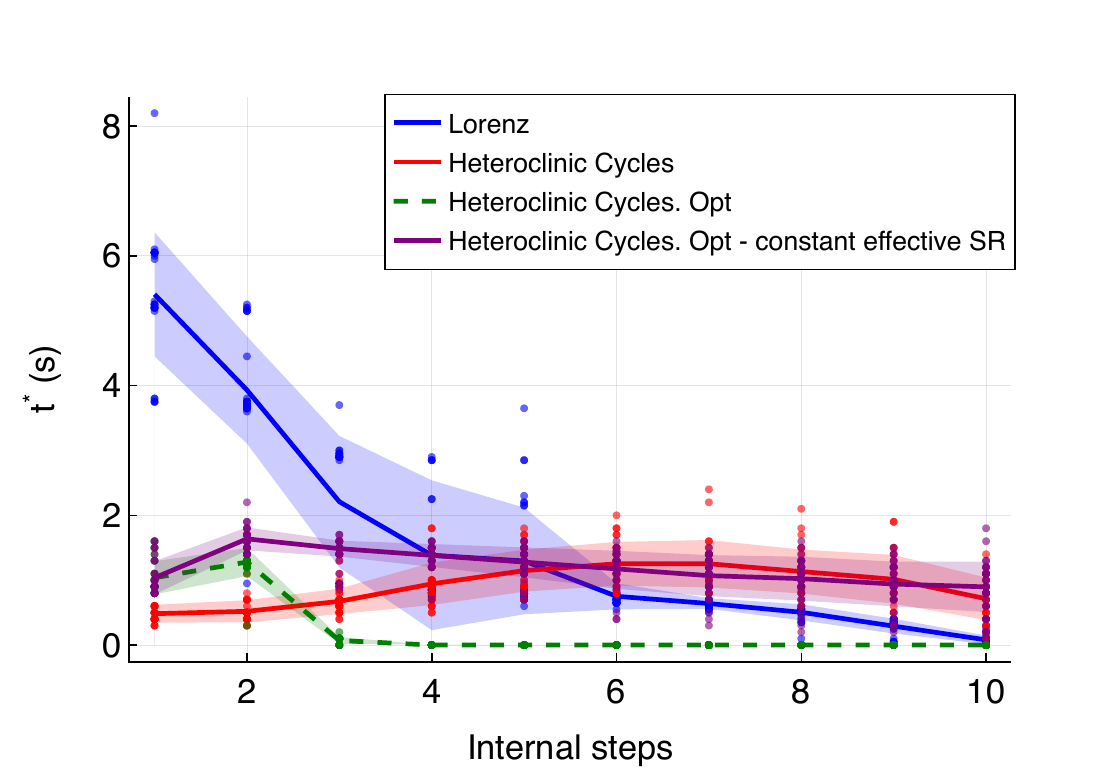}
\caption{Multi-step reservoir investigation. Mean valid times achieved by 30 multi-step reservoir instantiations as the number of internal steps varies on the Lorenz system (spectral radius $0.5$, input scaling $0.15$, regularization $0.000001$, others at baseline)(\textcolor{blue}{blue}), and the heteroclinic cycles regime under different parameter tunings. For un-tuned parameters (spectral radius $0.5$, input scaling $0.15$, regularization $0.000001$, others at baseline)(\textcolor{red}{red}), seven internal steps is a local optimum, significantly more than the usual single step. When the parameters are tuned for one internal step however (spectral radius $0.05$, input scaling $0.05$, regularization $0.0001$) (\textcolor{green}{green}) this optimum is only two steps. Subsequently, the performance drops off rapidly as the number of internal steps is increased. This may be due to the spectral radius being smaller in the optimized case than the untuned case such that the extra internal steps form an \textit{effective spectral radius} that is too small. It is also possible that the representations produced by a single step of the reservoir are already rich enough for the dynamical systems considered. When the spectral radius is adjusted from the tuned value such that the \textit{effective spectral radius}, defined as the spectral radius of $\textbf{A}^m$ where $\mathbf{A}$ is the internal reservoir connectivity matrix, and $m$ is the number of internal steps, is kept constant (at $0.05$), the rapid decrease in valid time is ameliorated (\textcolor{purple}{purple}). Maximum valid times achieved across these cases for the heteroclinic cycles regime do not differ greatly, suggesting optimization of the spectral radius (and other parameters, as conducted in this study) may be equivalent to optimizing the number of internal steps. Dots: individual reservoir instantiation valid time on single test span. Solid/Dashed lines: mean valid time across reservoirs. Shaded regions: one standard deviation across reservoirs.}
\label{supp_fig:multistep_reservoir}
\end{figure}

\begin{figure}[ht!]
\centering
\includegraphics[width=0.7\linewidth]{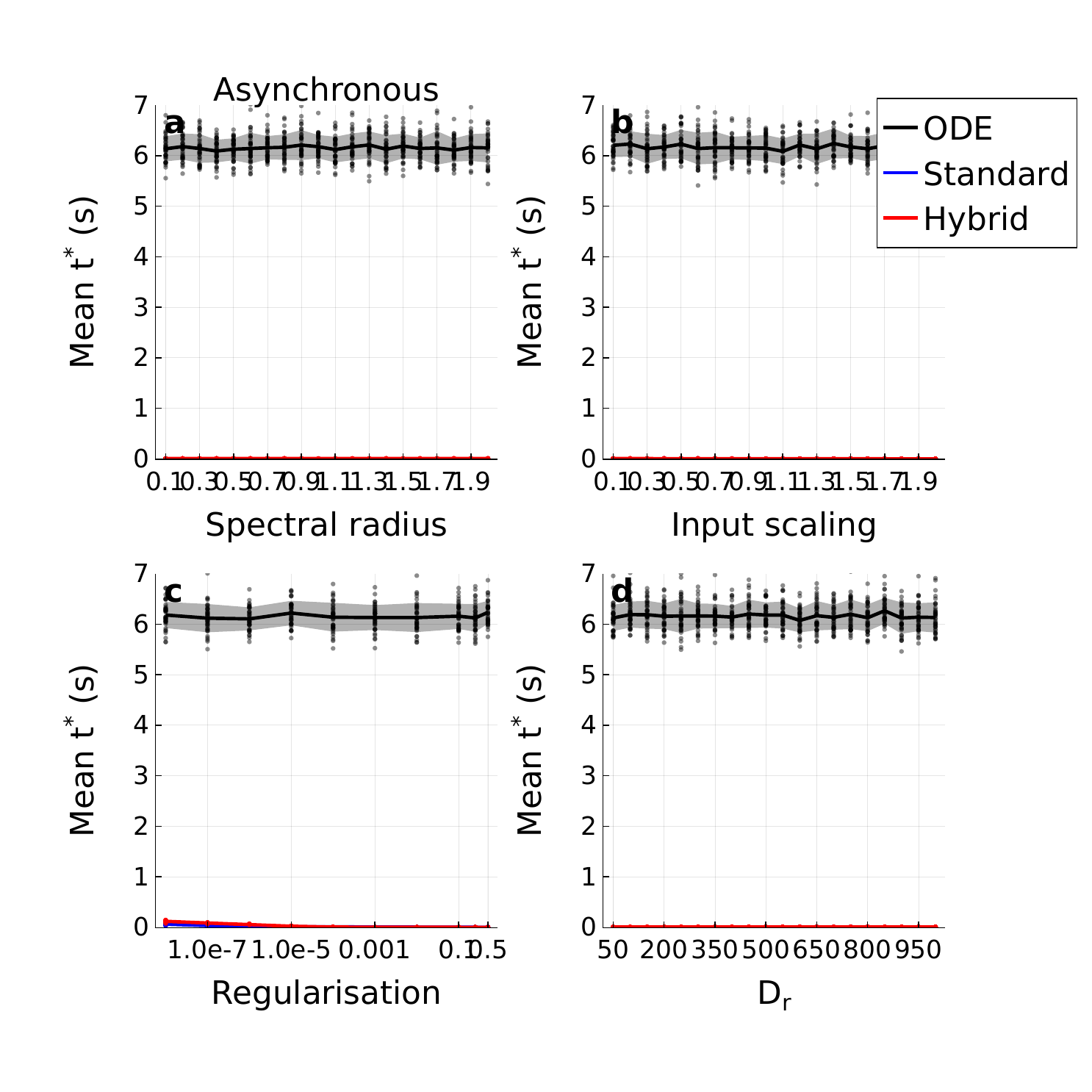}
\caption{Residual Physics task parameter sweeps evaluating the hybrid RC's prediction of NLON trajectories with missing dynamics in its expert model in a slow asynchronous regime. Mean valid time of the prediction of the hybrid RC (\textcolor{red}{red}), standard RC (\textcolor{blue}{blue}), and the base ODE model (black) as four different parameters are varied. Left to right, top to bottom - parameter varied: Spectral radius (\textbf{a}), Input scaling (\textbf{b}), Regularization (\textbf{c}), Reservoir size $D_r$ (\textbf{d}). Individual dots are individual reservoir/ODE instantiations (40), each representing the mean NMSE across 60 forecasts, (20 for each realization of a ground truth regime). Solid lines are the mean NMSE across the reservoir/ODE instantiations. Shaded regions are one standard deviation across reservoir/ODE instantiations.}
\label{supp_fig:slowAsync}
\end{figure}

\begin{figure}[ht!]
\centering
\includegraphics[width=\linewidth]{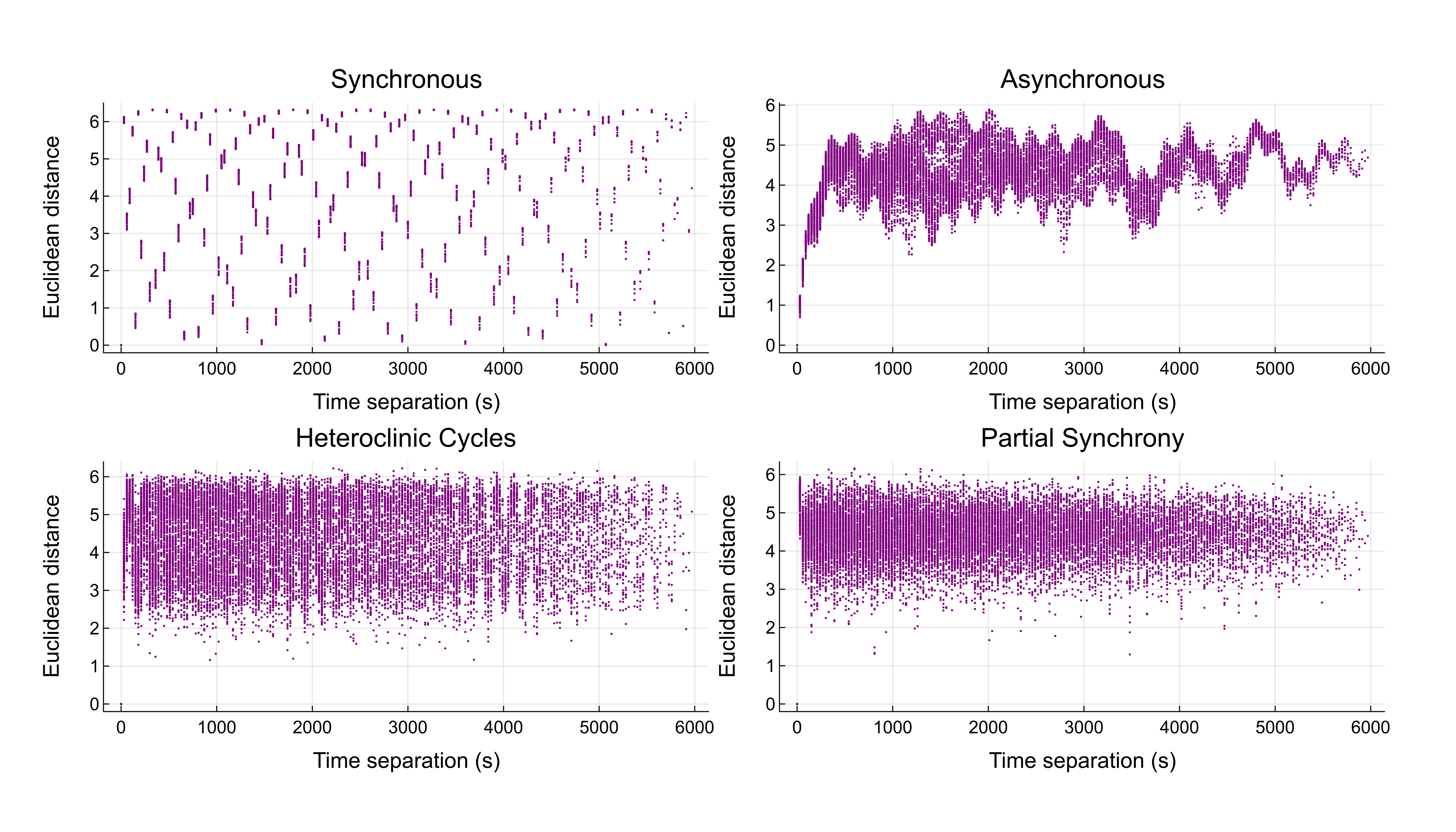}
\caption{Space-time separation plots of the ground truth trajectories of each dynamical regime of the bi-harmonic Kuramoto model from the residual physics task, with the slow asynchronous regime. These plot the pairwise separation between all pairs of points in the trajectory, with Euclidean distance on the y-axis and time on the x-axis. The slow asynchronous regime has a protracted region where for small time intervals, the Euclidean distance between points is close. This demonstrates how slow the trajectories are in comparison to the other regimes. We suggest that this is the cause of the training and prediction failures of both the standard and hybrid RCs on the slow asynchronous regime.}
\label{supp_fig:space_time_separation}
\end{figure}

\begin{figure}[ht!]
\centering
\includegraphics[width=0.7\linewidth]{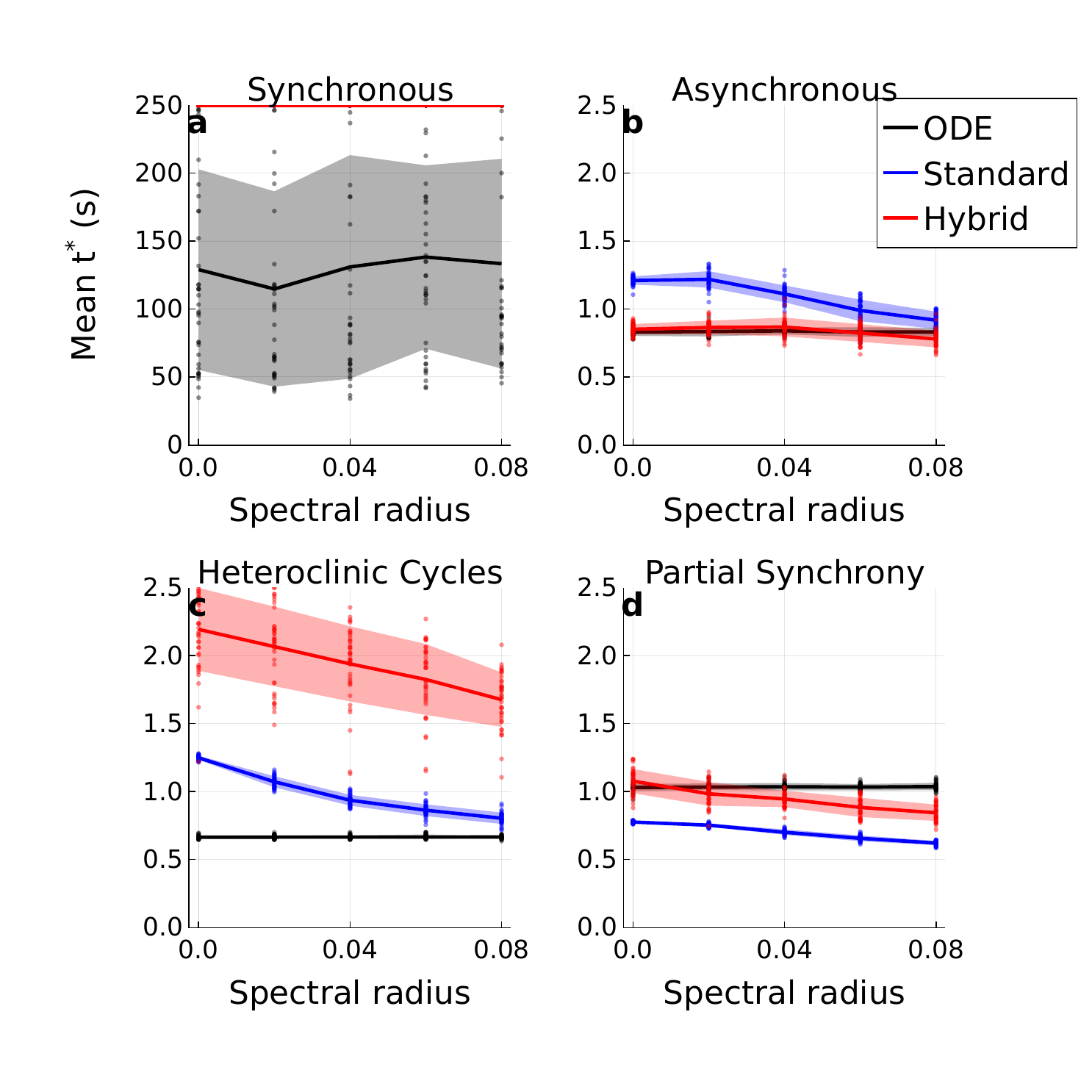}
\caption{Residual Physics task parameter sweeps evaluating the hybrid RC's prediction of NLON trajectories with missing dynamics in its expert model as the spectral radius of the reservoir is varied down to 0.0. Mean valid time of the prediction of the hybrid RC (\textcolor{red}{red}), standard RC (\textcolor{blue}{blue}), and the base ODE model (black) across the four dynamical regimes. Left to right, top to bottom - dynamical regime: Synchronous (\textbf{a}), Asynchronous (\textbf{b}), Heteroclinic Cycles (\textbf{c}), Partial Synchrony (\textbf{d}). Individual dots are individual reservoir/ODE instantiations (40), each representing the mean NMSE across 60 forecasts, (20 for each realization of a ground truth regime). Solid lines are the mean NMSE across the reservoir/ODE instantiations. Shaded regions are one standard deviation across reservoir/ODE instantiations.}
\label{supp_fig:lowSR}
\end{figure}

\begin{figure}[ht!]
\centering
\includegraphics[width=1.0\linewidth]{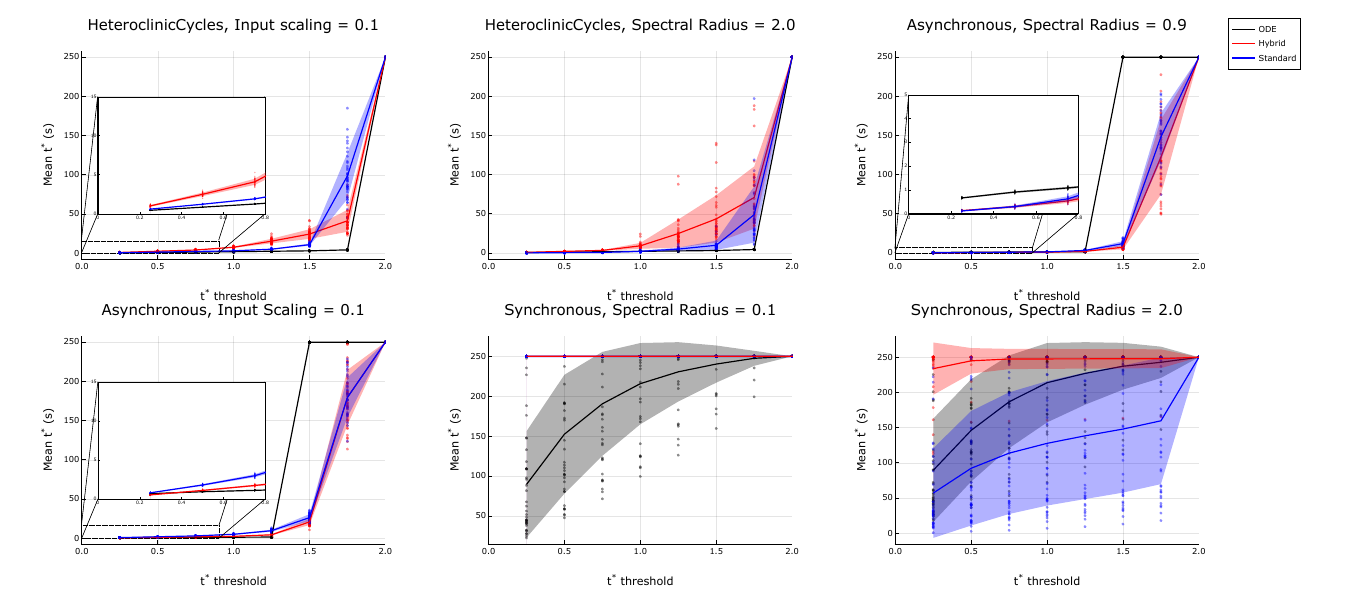}
\caption{Varying the threshold used in the valid time metric computation. Each subplot is at a particular parameter setting from the sweeps in Figure 9, for a given regime. Solid lines: mean mean valid time across tests and instantiations. Dots - mean valid time across tests per instantiation. Shaded region - one standard deviation over instantiations. A threshold of 2.0 is the maximum possible normalized mean square error, occurring when the ground truth state and model prediction are at opposite ends of the state space (20 dimensions). ``Accurate'' trajectories occur well below this, but this is subjective. A threshold of 0.4 is used in the paper, small variations about this do not significantly affect the ranking of the methods.}
\label{supp_fig:varyThresh}
\end{figure}
\clearpage
\subsection*{Code Implementation}
\textbf{Random Seeds} The random number generator used to instantiate each reservoir and ODE parameter settings, as well as the parameter error sampling was of Mersenne Twister type, with a seed set to $1234 + parameter\_index$, where the $parameter\_index$ ranged from $1$ to $20$ where the parameter sweep was across 20 points.

\end{document}